\documentclass[traditabstract]{aa} 
\usepackage{graphicx}
\usepackage{txfonts}
\begin{document}
   \title{A snapshot on galaxy evolution occurring in the Great Wall:}

   \subtitle{the role of Nurture at $z$=0}

   \author{Giuseppe Gavazzi \inst{1}, Mattia Fumagalli \inst{1},
          Olga Cucciati \inst{2}, Alessandro Boselli
	  \inst{2} 
          }
   \institute{Dipartimento di Fisica G. Occhialini, Universit\`a di Milano-Bicocca, Milano, Italy\\
             \email{giuseppe.gavazzi@mib.infn.it, mattia.fumagalli@mib.infn.it} 
	\and  Laboratoire d'Astrophysique de Marseille, Marseille, France \\ 
             \email{olga.cucciati@oamp.fr, alessandro.boselli@oamp.fr}
   	     }
              \date{Received January 28, 2010; accepted ...}

\abstract{
   With the aim of quantifying the contribution of the environment on the evolution of galaxies at $z$=0
   we have used the DR7 catalogue of the Sloan Digital Sky Survey (SDSS)
   to reconstruct the 3-D distribution of 4132 galaxies in 420 square degrees of the Coma supercluster, 
   containing two rich clusters (Coma and A1367), several groups, and many filamentary 
   structures belonging to the "Great Wall", at the approximate distance of 100 Mpc.
   At this distance the galaxy census is complete to $M_i=-17.5$ mag, i.e. $\sim$ 4 mag fainter than $M^*$.
   The morphological classification of galaxies 
   into early- (ellipticals) and late-types (spirals) was carried out by inspection 
   of individual SDSS images and spectra.
   The density around each galaxies was determined in cylinders of 1 Mpc radius
   and 1000 km s$^{-1}$ half length. 
   The color-luminosity relation was derived for galaxies 
   in bins morphological type and in four thresholds of galaxy density-contrast, ranging from  
   $\delta_{1,1000} \leq 0$ (UL = the cosmic web); 
   $0 < \delta_{1,1000} \leq 4$  (L = the loose groups); $4 < \delta_{1,1000} \leq 20$ (H = the large groups
   and the cluster's outskirts) and  $\delta_{1,1000} > 20$ (UH = the cluster's cores).
   The fraction of early-type galaxies increases with the log of the over-density.
   A well defined "red sequence" composed of early-type galaxies exists in all environments 
   at high luminosity, but it lacks of low luminosity (dwarf) galaxies in the lowest density environment.
   Conversely low luminosity isolated galaxies are predominantly of late-type. In other words
   the low luminosity end of the distribution is dominated by red dE galaxies in clusters and groups
   and by dwarf blue amorphous systems in the lowest density regions.
   At $z$=0 we find evidence for strong evolution induced by the environment (Nurture).
   Transformations take place mostly at low luminosity when  
   star forming dwarf galaxies inhabiting low density environments migrate into 
   amorphous passive dwarf ellipticals in their infall into denser regions. 
   The mechanism involves suppression of the star formation due to gas stripping,
   without significant mass growth, as proposed by Boselli et al. (2008a). 
   This process is more efficient and fast in ambients of increasing density.
   In the highest density environments (around clusters) the truncation of the star formation 
   happens fast enough (few 100 Myr) to produce the signature of post-star-burst in galaxy spectra.
   PSB galaxies, that 
   are in fact found significantly clustered around the largest dynamical units,
   represent the remnants of star forming isolated galaxies 
   that had their star formation violently suppressed during their infall in clusters in the last 0.5-1.5 Gyrs, 
   and the progenitors of future dEs.
   }


%
\authorrunning{Gavazzi et al.}
\titlerunning{A snapshot on galaxy evolution in the Great Wall} 
\maketitle

\section{Introduction}

In 1986 Valerie de Lapparent, Margaret Geller and John Huchra announced 
their "Slice of the Universe", a strip of 1100 galaxy optical redshifts from the CFA
redshift survey, going through the Coma cluster (de Lapparent et al. 1986). 
A spectacular connected filament of galaxies popped out from their data, 
later nicknamed the "Great Wall" by Ramella et al.
 (1992, who more than doubled the number of redshift measurements) 
running through the spring sky at an approximate distance of 100 Mpc 
(adopting the modern value of $Ho$=73 km s$^{-1}$ Mpc$^{-1}$, as in the remaining of this paper), 
and spanning a similar length.
Its narrowness in the redshift space, bracketing large voids, was shocking for the epoch, for an universe  
supposedly "homogeneous and isotropic" on that scale.\\
Hundreds individual galaxies in the region $11.5^h<RA<13.5^h;+18^o<Dec<32^o$ 
that covers the majority of the the Coma Supercluster (CS) were in the mean time being
surveyed at 21 cm (e.g. Chincarini et al. 1983; Gavazzi 1987, 1989, Scodeggio \& Gavazzi 1993 and references therein) providing additional radio redshifts. 
Gavazzi et al. (1999), following the pioneering work of Gregory \& Thompson (1978),
highlighted the 3-D structure of the CS, containing 2 rich clusters: Coma (A1656) and A1367,
several minor dynamical units (groups) and a multitude of galaxies belonging to the
filaments with a small velocity dispersion.
Being at constant distance from us, the CS region offered a 
unique opportunity for comparing directly the 
properties of galaxies in very different environmental conditions.
As the spread in redshift is minimal in the CS (let alone the fingers of God of the two clusters) 
the high density contrast with foreground and background structures 
(large volumes in the foreground of the CS
are completely devoid of galaxies) makes it
quite straightforward to identify a huge slab of galaxies at approximately constant
distance ($D$=96 Mpc, $\mu=-34.9$ mag corresponding to $cz\sim $7000 km s$^{-1}$), making it possible to compare 
galaxies in extreme environments free from distance biases.\\
By 2003, when the data from the Coma region were providing the skeleton for the web-based
Goldmine database (Gavazzi et al. 2003), 
the 1127 CS galaxies included in the CGCG catalogue (Zwicky et al. 1961-1968) had all a measured redshift.
Due to the shallowness of the CGCG catalogue, however, 
only giant galaxies ($m_p$ $\leq$ 15.7 mag corresponds to 
$M_p$ $\leq$ -19.3 at the distance of CS) were represented, while dwarfs,  
that meanwhile were discovered to dominate the galaxy luminosity function, were neglected.\\ 
The idea of undertaking a deeper photometric survey of the CS region surpassing by at least 2 mag the CGCG 
was considered and soon abandoned because it would have been prohibitive for medium size telescopes 
equipped with CCDs not larger than 10 arcmin on each side, as available in the late '90s, to cover
such a large stretch of the sky. Let alone the spectroscopy: 
for $r<18$ mag there are about 150 galaxies per square deg. (Heidt et al. 2003; Yasuda et al. 2007), 
so that in 420 $\rm deg^2$ it would have been necessary to take 65000 spectra for sorting out 
the objects in the velocity range $4000<cz<9500$ km s$^{-1}$ appropriate for the CS 
(that a posteriori we know to sum only to approximately 4000).\\
With the DR7 data release of SDSS (2009) this dream became a reality:
not only can we rely on a five color ($u,g,r,i,z$) photometric survey 
(with limiting sensitivity of 22.2  mag in the $g$ and $r$ bands
and limiting surface brightness of $\mu_g \sim 25.5 \rm ~mag~ arcsec^{-2}$), but also 
on the spectroscopy of (nominally) all galaxies brighter than $r<17.77$ mag\footnote{The success rate of redshift
determinations from SDSS spectra is reported to be $>99\%$ by Stoughton et al. (2002).} (with
effective surface brightness $\mu_e<24.5 \rm ~mag~ arcsec^{-2}$).\\ 
The present "near-field cosmology" project makes use of the photometry 
and the spectroscopy provided by SDSS much in the way astronomers use 
telescopes in "service observing" mode, but counting on an almost unlimited amount of observing time.
By saying so we wish to emphasize that,
opposite to most SDSS statistical analyzes that are based on such large number of objects that
cannot be inspected individually,
the number of galaxies ($\sim$ 4000) in the present study is large enough to ensure statistical
significance, but not too large to prevent individual inspection of all images and spectra.
The Coma supercluster constitutes the ideal bridge between the near-field and the high-$z$ cosmology.\\ 
Several modern studies based on hundreds of thousands galaxies from the SDSS in the local universe
(usually $0.05<z<0.2$) address environmental issues by quantifying the dependence of galaxy structural 
and photometric parameters from the local galaxy density. The density around each galaxy is computed 
counting galaxies in cylinders of given projected radius and depth in the redshift space. 
Most studies adopt 2 Mpc and 500 $\rm km s^{-1}$ (e.g. Kauffmann et al. 2004) or
1 Mpc and 1000 $\rm km s^{-1}$ (e.g. Hogg et al. 2004), as a compromise for detecting structures intermediate between 
large groups and small clusters. 
It is not easy however to visualize what type of structures are found by these automatic algorithms.
The present study gives the opportunity to reciprocally calibrate structures that are seen individually
in the CS with those that are detected by statistical means. \\
It is well know that the SDSS pipelines have not been designed for extracting parameters 
of extended nearby galaxies, which results in problems in creating reliable 
low-redshift catalogs, due to:
1) "shredding" of large galaxies in multiple pieces leading to wrong magnitude determinations.
2) slight (20\%) underestimate of the brightness of galaxies larger than $r_{50}\sim 20"$ due to
the way the photometric pipeline determines the median sky, 
biasing the sky brightness near large extended galaxies, 
as reported by Mandelbaum et al. (2005), Lauer et al. (2007), Bernardi et al. (2007), 
and Lisker et al. (2007).
3) The SDSS spectroscopy is meant to be complete down to $r<17.77$ mag (Strauss et al. 2002). However several
sources of incompleteness affect it, primarily due to conflicts in the fiber placement 
(they cannot be placed more closely than 55") and because the correlation 
between galaxy surface brightness and luminosity biases against faint objects.
All these sources of incompleteness have been considered by Blanton et al. (2005a,b).\\
In the Coma supercluster the problems connected with the incompleteness of SDSS are less severe than elsewhere.
Even the largest of its galaxies have diameters $<2-3$ arcmin, making the "shredding" problem much less
serious than in nearer surveys (e.g. Virgo). The brightest galaxies are well cataloged 
and their photometric parameters can be re-determined manually using the SDSS navigator.
In principle the spectroscopy of several bright galaxies in the CS can be obtained from the literature beside the SDSS.
Moreover even the smallest galaxies belonging to the CS
have sizes in excess of 10 arcsec. This allows their classification by visual inspection
on SDSS plates, independent from, but consistent with the method of visual morphology undertaken by
the the Galaxy Zoo project (Lintott et al. 2008).\\
In conclusion the SDSS offers a magnificent tool to study a sizable stretch of the local universe,
providing a representative census of galaxies at $z=0$, their color, morphology, luminosity and spectral classification and 
allowing us to compare these quantities free from distance biases
in a variety of environmental conditions, from the center of two rich clusters 
to the sparse filamentary regions where the local galaxy density is $\sim$ 100 times lower.\\
The Coma supercluster offers a unique photographic snapshot of the evolutionary status of galaxies
at $z$=0 in a variety of environmental conditions that can be used to constrain the
mechanisms that drive the evolution of galaxies at the present cosmological epoch 
(see a review in Boselli \& Gavazzi 2006).
It provides a complementary view to
the  many statistical determinations that are currently obtained using the unique data-set provided by SDSS
(e.g. Kauffmann et al. 2004; Balogh et al. 2004a,b; Blanton et al. 2005a,b,c; Martinez \&  
Muriel 2006; Blanton \& Berlind 2007, Haines et al. 2007, just to mention a few,
culminating with the review of the physical properties of nearby galaxies and their environment
by Blanton \& Moustakas 2009). 
Combined with studies of galaxy evolution at higher redshift
(e.g. Scarlata et al. 2007 and Cassata et al. 2007 for $0<z<1$; Faber et al. 2007 for $1<z<2$;
Fontana et al. 2006 for $1<z<4$) 
it will shed light on the mechanisms and processes that contribute to the evolution of galaxies 
in the cosmological context. 

\section{Data}

\subsection{Photometry}

The SDSS DR7 spectroscopic database was searched in the window 
$11.5^h<RA ~(J2000) <13.5^h;+18^o<Dec ~(J2000)<32^o$
($\sim 420$ $\rm deg^2$) for all galaxies 
with a measured redshift in the interval $4000<cz<9500$ km s$^{-1}$
\footnote{The interval $4000<cz<9500$ km s$^{-1}$ was quite arbitrarily selected in order
to include the whole "finger of God" associated to the Coma cluster.}.
We obtained $\sim$ 4000 targets. For each one we took coordinates, 
$u,g,r,i,z$ Petrosian magnitudes (AB system) and the redshift.\\
To fill-in the incompleteness for luminous galaxies (Blanton et al. 2005a,b) we added 101 CGCG
galaxies with redshift known from Goldmine that were not 
included in the SDSS spectral database.  
Also for these objects we took the $u,g,r,i,z$ Petrosian magnitudes using the SDSS DR7 navigation tool,
that provides reliable magnitudes.\\
Additional galaxies that might have been missed in
SDSS spectroscopic catalogue mainly due to fiber conflict were
searched in NED within the limits of the CS window ($\Delta \alpha; \Delta \delta; \Delta cz$).
We obtained only 160 of these serendipitous targets. Also for them we took 
the $u,g,r,i,z$ magnitudes using the SDSS navigator tool and among them we selected those meeting 
the condition $r\leq17.77$ 
(that matches the selection criterion of the SDSS spectral catalogue).
Only 76 such objects were left, emphasizing the high completeness of the SDSS spectroscopic set. 
As expected from the highest degree of crowding, a significant fraction (50\%) of the NED data 
concentrates on the two clusters, with the remaining 50\% objects spread 
around the whole supercluster. \\
Some galaxies in the SDSS list have their spectrum taken in a off-nuclear position.
Their given magnitudes are often not representative of the whole galaxy.
For these we re-measured the $u,g,r,i,z$ magnitudes at the central position using the SDSS navigator, 
and we assigned this set of magnitudes 
to the original SDSS redshift, discarding the spectral information. 
Furthermore we cleaned the SDSS list from double entries: e.g. galaxies with multiple spectra.
We remained with 4132 galaxies (3955 from SDSS; 101 from CGCG; 76 from NED) with redshift and photometric parameters.\\
We did not apply corrections for surface brightness because they are negligible at the present 
sensitivity limit (Blanton et al. 2005b). 
Even so, with the contribution of faint objects from NED the catalogue
should be nearly (94\%) complete down to $r=17.77$ ($i \sim 17.5$) according to Strauss et al. (2002).\\
A visual morphological classification aimed at disentangling early 
(dE, dS0, E, S0, S0a) from late-type (Sa to Sd, dIrr, BCD and S...) galaxies
was carryed out using the SDSS on-line explorer. 
Since the present analysis is based solely on a two bin classification scheme (early vs. late)
we considered sufficient that only one of us (GG) carried out the inspection of all 4132 galaxies. \\
Giant galaxies (E, S0, S0a, Sa, Sab, Sb, Sbc, Sc, Sd) with mild inclination 
were classified in steps of half Hubble class 
on a purely morphological basis, irrespective of their spectral classification. 
For 224 of them we can rely on the independent classification from Goldmine. 
The agreement is within $\sigma_{type}$ = 0.7 Hubble class.
Instead, for high inclination objects, whose possible spiral structure could be hidden
and whose color is reddened by internal absorption, our classification relied also on the inspection to
the spectra: we separated S... from S0s according to the presence/absence of emission lines.
Eventually, the fraction of emission line objects turned out to be negligible (5 \%) among E and S0 galaxies, 
it increased to 30 \% among S0a, and to 70 \% among S... objects.
S0a are dubious transitional objects, as they 
are separated from S0s only for having a slighltly more evident disk component. They were 
grouped among early type galaxies.\\
Among dwarf galaxies, blue compact objects (BCD) were classified according to their 
amorphous compact morphology, blue color and strong emission lines.
dIrr are blue, low surface brightness, emission line objects.
Amorphous early-type dwarfs are red, absorption line systems and were
classified as dE or dS0 according to the existence of lenticular components.
It should be stressed that BCDs, dEs and S..., often have similar structureless morphology.
Their difference is mainly dictated by their color and by the presence/absence of strong 
emission lines in their spectra (see also Fig.\ref{fig12}). \\

\subsection{Spectroscopy}
\label{spec}

All 4132 galaxies in the present sample have a redshift measurement, but only for 3955 
a spectrum was taken in the SDSS (in the central 3" of the nucleus). 
For the remaining 177 we searched NED and Goldmine for available 
spectra and found only 11 (taken in the drift-scan mode, including the nucleus).\\
For the 3955 galaxies with a nuclear spectrum we obtained from the SDSS spectroscopic database
the intensity and the equivalent width of the principal emission and absorption lines 
(from H$\epsilon$ $\lambda$ 3970
to [SII] $\lambda$ 6731) along with their signal-to-noise ratios.\\
Among spectra whose Balmer lines have signal-to-noise ratio $>$ 5
we identify 53 galaxies with Post-Star-Burst signature (PSB or k+a)
adopting the criterion of Poggianti et al. 2004 (inherited from Dressler et al. 1999)
(see however Quintero et al. 2004)
that no emission lines should be present in the spectra and:
\begin{equation}
\centering
EW(H\delta)>3  ~~\AA
\label{eqPSB}
\end{equation}
(adopting the convention that lines in absorption have positive EW).
The PSB galaxies are analyzed in Sect. \ref{PSB}.

\section {Analysis}

\subsection{Quantifying the local density.}

\begin{figure*}
 \centering
 \includegraphics[width=13.0cm]{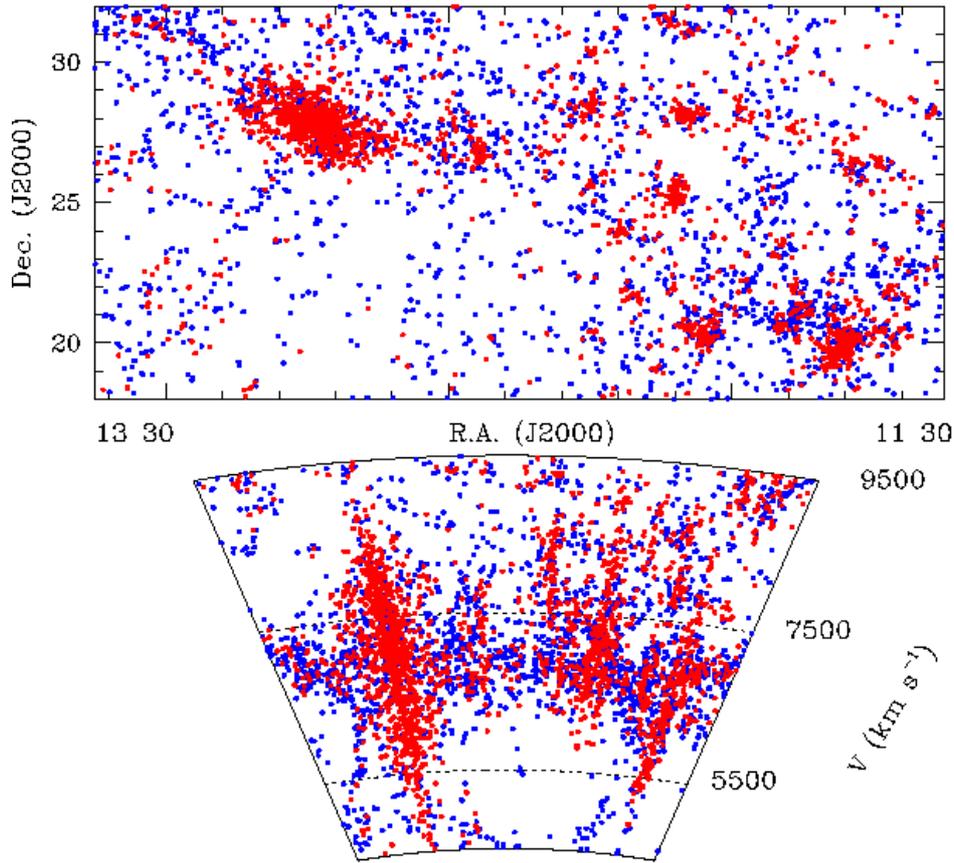}
 \caption{Celestial distribution of 4132 galaxies from SDSS in the Coma supercluster
 $11.5^h<RA<13.5^h;18^o<Dec<32^o$; $4000<zc<9500$ $\rm km s^{-1}$ (top panel) and their wedge diagram (bottom panel)
 Objects are color coded according to their morphological type 
 (E+S0+S0a are red; Spirals, BCD, Pec, S... are blue)
 \label{fig1}}
 \end{figure*}
  \begin{figure*}
  \centering
  \includegraphics[width=13.0cm]{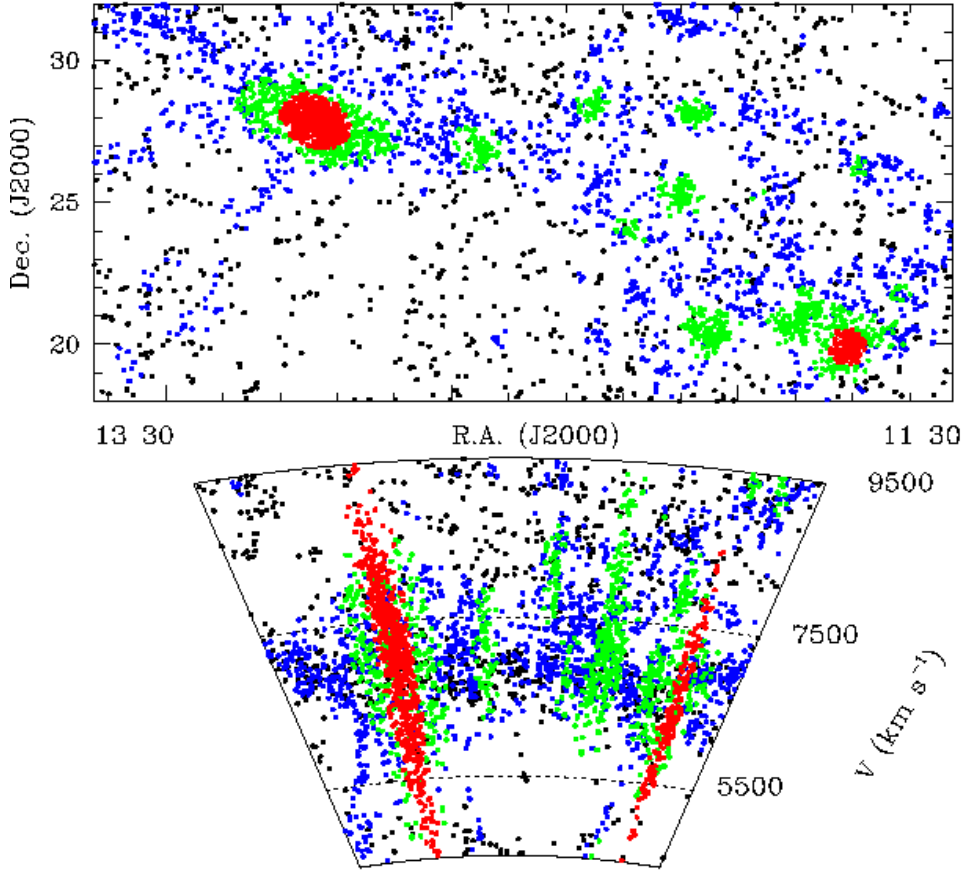}
  \caption{Same as Fig. \ref{fig1} except that
  objects are color coded according to their over-density (see Section 3.1) as follows: UH ($\delta_{1,1000}>20$) are red; 
  H ($4<\delta_{1,1000}\leq20$) are green; L ($0<\delta_{1,1000}\leq4$) are blue; UL ($\delta_{1,1000}\leq0$) are black.
  \label{fig2}}
  \end{figure*}
  \begin{table*}
  \centering
  \caption{Mean coordinates, $cz$, velocity dispersion, density and number of elements
 of clusters and groups in the H density bin, with $N \geq 20 $ members.\label{tbl-1}}
\begin{tabular}{lrcccccc}
\hline
\hline
 & NGC & $<RA>$ & $<Dec>$ & $<cz>$ &  $\sigma_v$ & $\delta_{1,1000}^{max}$& N \\
 & & deg & deg & $\rm km s^{-1}$ &  $\rm km s^{-1}$ &  \\ 
\hline
Coma   & 4889   &   194.157  & 27.87 & 6972 & 940 & 60.4& 750   \\
A1367  & 3842   &   176.243  & 19.93 & 6494 & 762 & 27.7& 220   \\
grp-1  & 4065   &   181.069  & 20.46 & 7055 & 401 & 16.7& 155	\\   
grp-2  & 3937   &   178.039  & 20.71 & 6659 & 294 & 12.0& 83	\\   
grp-3  & 128-34 &   182.027  & 25.30 & 6707 & 349 &  9.0& 78	\\   
grp-4  & 4555   &   189.024  & 26.85 & 7083 & 415 &  7.6& 64	\\   
grp-5  & 4104   &   181.556  & 28.14 & 8410 & 450 &  8.9& 62	\\   
grp-6  & 4295   &   185.054  & 28.47 & 8069 & 320 &  6.6& 51	\\   
grp-7  & 3910   &   177.616  & 21.36 & 7874 & 207 & 11.8& 43	\\   
grp-8  & 4213   &   183.787  & 24.01 & 6753 & 268 &  4.7& 26	\\   
\hline
\end{tabular}
\\
\end{table*}
\begin{table*}
\centering
\caption{Percentage of galaxies in the various density bins (S0a are included among early-type).\label{tbl-2}}
\begin{tabular}{lcccccccc}
\hline
\hline
             & 	$\delta_{1,1000}$  & All  & Red & $\rm Lum/Faint_{Red}$& Blue	 & Early & Late & PSB \\
             &                     & N	  &  \% &   		      & \%   & \%    &  \%  &  \% \\
\hline
UL           &	$\leq$0   	   & 835  & 22  & 82		      & 78   & 18    & 82   & 0.5 \\
L            &	0-4       	   & 1476 & 39  & 56		      & 61   & 35    & 65   & 0.8 \\
H            &	4-20      	   & 1032 & 63  & 47		      & 37   & 61    & 39   & 2.4 \\
UH           &	$>$20     	   & 789  & 82  & 32		      & 18   & 84    & 16   & 7.3 \\
\hline
All          &	          	   & 4132 &  50 & 47		      & 50   & 47    & 53   & 2.6 \\
\hline
\end{tabular}
\\
\end{table*}
\begin{figure*}[!t]
\centering
\includegraphics[width=10.0cm]{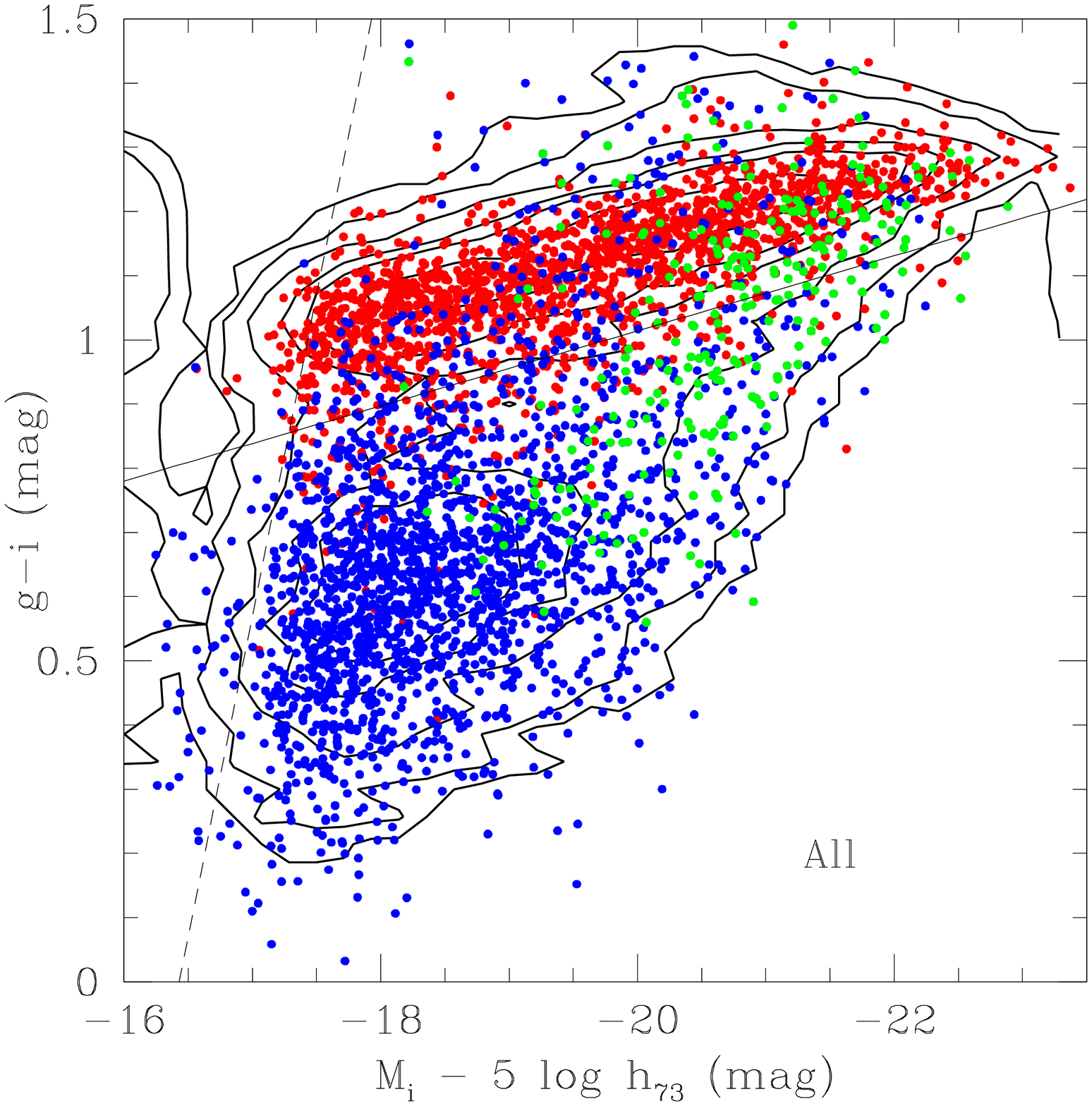}\\
\includegraphics[width=5.0cm]{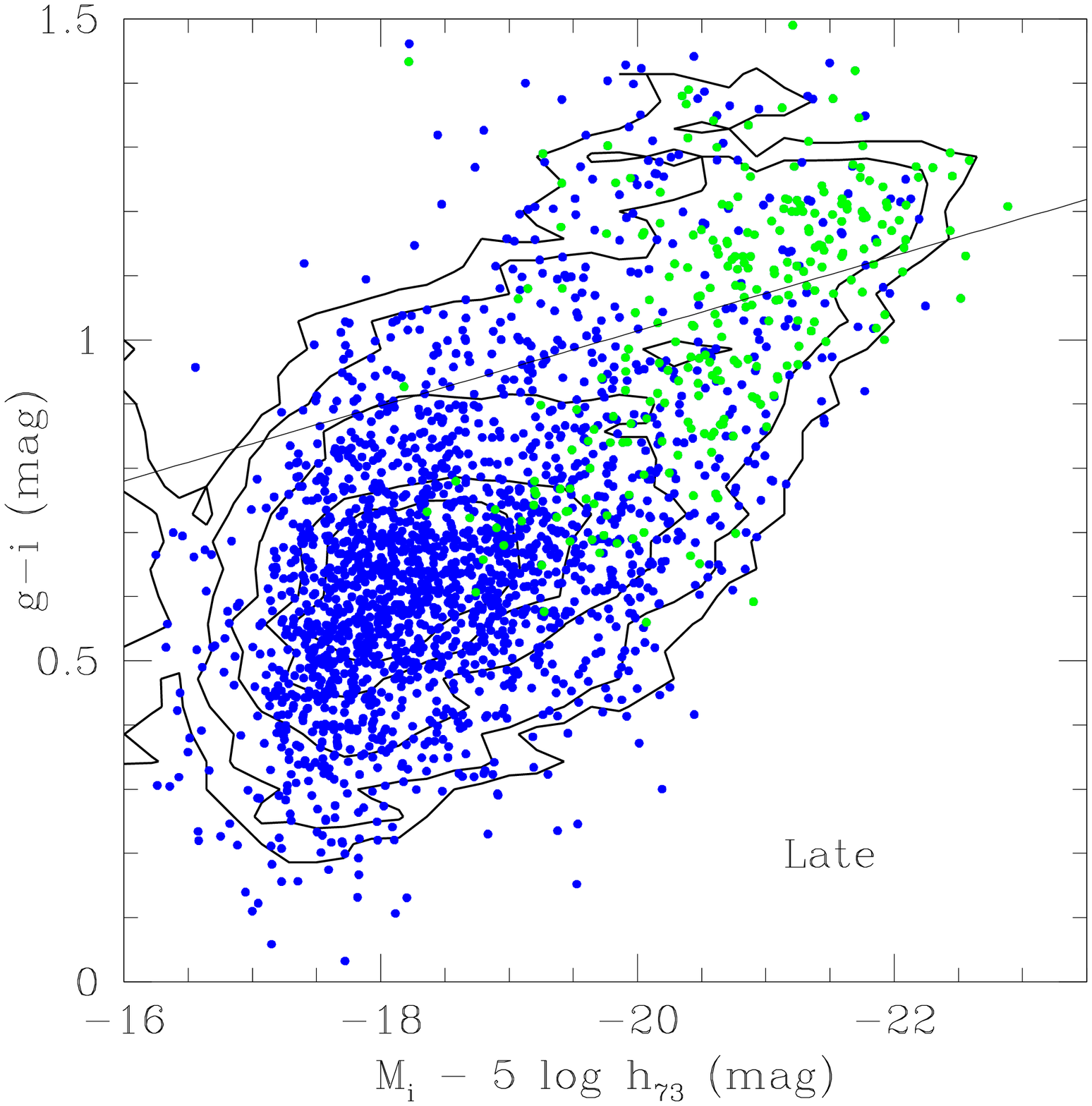}
\includegraphics[width=5.0cm]{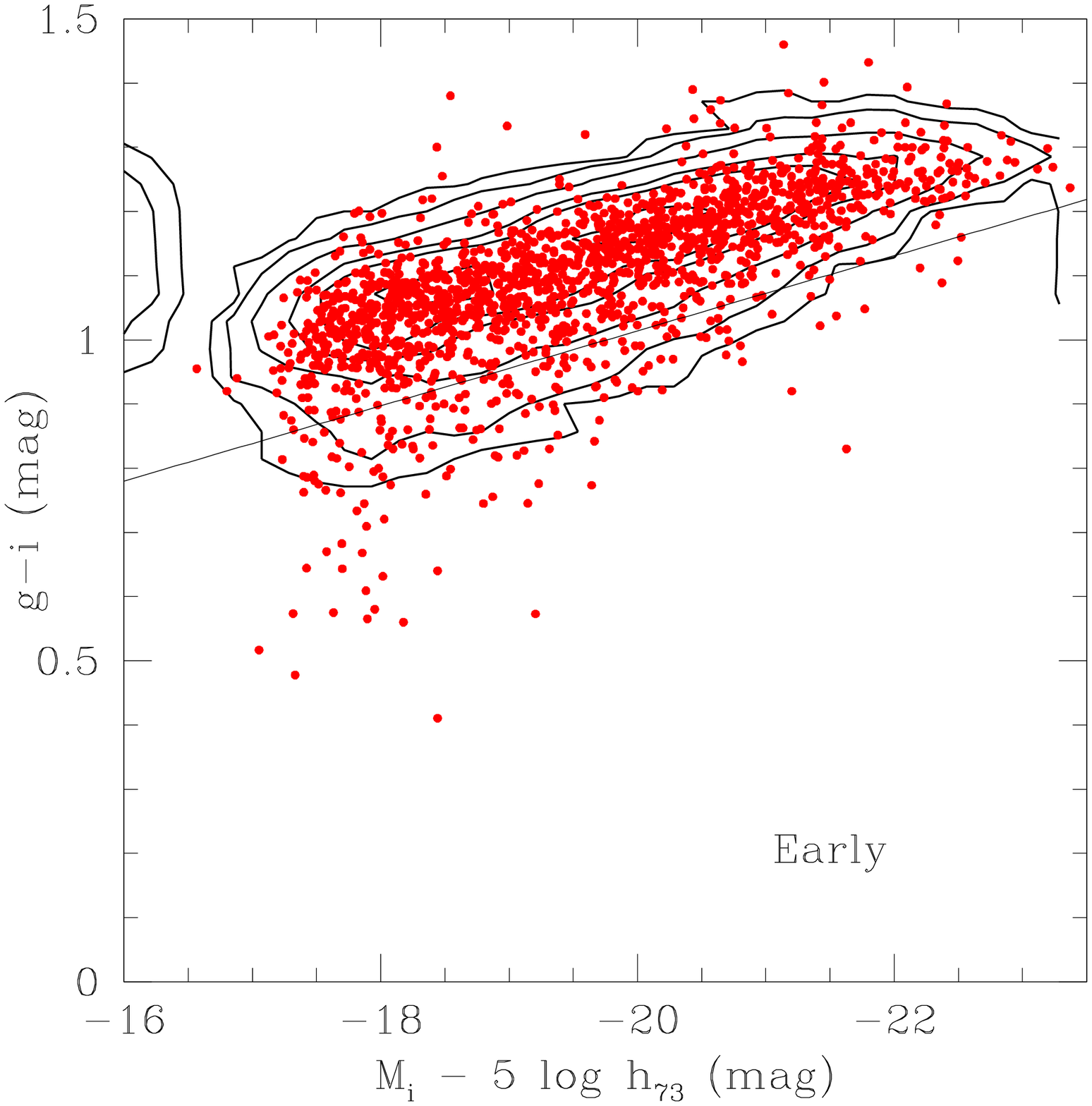}
\caption{$g-i$ color versus $i$-band absolute magnitude relation of all galaxies in the CS coded according 
to Hubble type: red = early-type galaxies (dE-E-S0-S0a); blue = disk galaxies (Sbc-Im-BCD); green = bulge galaxies 
 (Sa-Sb): all galaxies (top); late-type from Sa to Im-BCD (bottom left); early type (bottom right). Contours of equal density are given. 
The continuum line $g-i = -0.0585 * (M_i +16)+0.78$ represents the empirical separation between the red-sequence and the remaining
galaxies (see note 3). The dashed line illustrates the effect of the limiting magnitude $r$=17.77 of the spectroscopic SDSS database, 
combined with the color of the faintest E galaxies $g-r ~\sim~0.70$ mag. 
\label{figure3}}
\end{figure*}
Figure \ref{fig1} shows the modern version of the "Slice in the Universe" 
obtained with the SDSS data.
The celestial distribution and the wedge diagram are given for the 4132 galaxies in the
present study, coded according to their morphological classification (red: early-type; blue: late-type).
The two rich clusters, Coma and Abell 1367 and some minor concentrations dominated by early-type galaxies are apparent,
clearly indicating the well known correlation between the morphology of galaxies and their 
surrounding environment (e.g. Dressler 1980; Blanton \& Berlind 2007).\\
Various approaches have been used to characterize the environment of galaxies.  
One possibility is to recognize groups and clusters, compute their virial mass from 
the velocity dispersion and assign a density to all the member galaxies. 
Another possibility consists in evaluating a local density parameter for each galaxy, 
by counting the number of galaxies in cylinders of given radius and half-length centered on each galaxy.
This is the typical strategy of large surveys, for example followed by Kauffmann 
et al. (2004) ($2$ Mpc, $500~$$\rm km s^{-1}$), 
and Hogg et al. (2004) ($1h^{-1}$Mpc, $8 h^{-1}$Mpc $\sim$ 1000 $\rm km s^{-1}$).\\
The typical half-lengths and radii used in the literature appear immediately inadequate to 
describe a region as the one under study, featuring two rich clusters with a velocity 
dispersion as large as $\sim 1000$ $\rm km s^{-1}$. 
For instance, if we chose a cylinder of 500 or 1000 $\rm km s^{-1}$ depth, 
we would introduce a severe bias because galaxies with the largest velocity dispersion 
(i.e. the ones belonging to the extremities of the "Fingers of God" of the two clusters) would come out 
with a much lower density than the real density of the cluster they belong to, as noted by Lee et al. (2009). 
Conversely, if we  used cylinders of 2000-3000 $\rm km s^{-1}$ length to 
evaluate the density around a relatively isolated object
located far away from the two clusters, its density would came out overestimated because
background and foreground galaxies not physically associated with it would 
wrongly contribute to the density.\\
In order to comply with criteria  that are generally adopted in the literature 
(namely $1$ Mpc, $1000~$$\rm km s^{-1}$), so that the results from this work could be easily compared with 
other reference papers, but also keeping an eye on the physical structures that constitute the CS,  
we decided to evaluate the density adopting the following compromise strategy.
First we compress the "Fingers of God" of Coma and Abell 1367 in the redshift space by assigning to
galaxies belonging to the "Fingers of God" a velocity
equal to the average velocity of the cluster, plus or minus a random Gaussian distributed 
$\Delta V$ comparable to the transverse size of the cluster on the plane of the sky, assuming that
clusters have approximately a spherical shape in 3-D. (We assume that the transverse size is 2 deg and 1 deg
for Coma and A1367, corresponding to 248 and 124 $\rm km s^{-1}$ respectively). \\
Then we parameterize the environment surrounding each galaxy using the 3-D density contrast computed as: 
$$\delta_{1,1000} = \frac{\rho-<\rho>}{<\rho>}$$
where $\rho$ is the local number density and $<\rho>$ = 0.05 gal $(h^{-1}~$Mpc$)^{-3}$ 
represents the mean number density measured in the whole region. 
The local number density $\rho$ around each galaxy is computed within a cylinder 
with  1 $h^{-1}$Mpc radius and 1000 $\rm km s^{-1}$  half-length,  which is large enough
to comprise the dispersion of small groups and the newly compressed Finger of Gods 
of Coma and Abell 1367. To take into account boundary effects (the cylinders centered on galaxies near the edges of 
our sample will partially fall outside the studied volume), for each galaxy we divide $\rho$ by the fraction 
of the volume of the cylinder that fall inside the survey borders. \\
We divide the sample in four over-density bins, chosen in order to highlight physically different
environments of increasing level of aggregation (see Fig.\ref{fig2}):
The UltraLow density bin (UL: $\delta_{1,1000}\leq 0$) describes the underlying cosmic web;
the Low density bin (L: $0 < \delta_{1,1000} \leq 4$) comprises the filaments in the Great Wall and the
loose groups; the High density bin (H: $4 < \delta_{1,1000} \leq 20$) include
the cluster outskirts and the significant groups; the UltraHigh density bin (UH: $\delta_{1,1000} > 20$) 
corresponds to the cores of the clusters.\\
We would like to emphasize that the Coma Supercluster is by definition over-dense with respect to the mean Universe. 
Hogg et al. (2004) find a general density of 0.006 gal $(h^{-1}~$Mpc$)^{-3}$ for $M_{i} < -19.5$. 
Our sample has a mean density of 0.05 gal $(h^{-1}~$Mpc$)^{-3}$ for galaxies with $M_{i} < -17.5$.
Adopting $M_{i} < -19.5$ the mean density in the Coma Supercluster becomes 0.019 gal $(h^{-1}~$Mpc$)^{-3}$,
thus we conclude that we are observing a region 
three times more dense, on average, than the mean density of the Universe.
However if from the CS sample we subtract 916 galaxies in the densest bins UH 
and H, contributed to by the Coma cluster alone,
the mean density drops to 0.015 gal/$h^{-1}$Mpc$^{3}$, i.e. 2.5 times over-dense than average.\\
Figure \ref{fig2} shows the celestial distribution and the wedge diagram for the 4132 galaxies in the
present study, coded according to the 4 over-density bins.
Combined with Fig. \ref{fig1}, Fig. \ref{fig2} indicates that the majority of 
early-type galaxies belongs to the UH and H bins,
while the majority of late-type galaxies are in the L and UL bin.\\ 
The parameters of the two clusters and 8 most populated groups identified within the H density 
bin are listed in Table \ref{tbl-1}. Some of these groups are known to contain radiogalaxies with complex 
morphology, suggestive of ram-pressure bending. 
The wide angle tail (WAT) radiogalaxy associated with N4061 is in group 1 (NGC 4065) (Jaffe et al. 1986).
N4061 was also detected in X-ray by (Doe et al. 1995) implying a central IGM density of
$4.9 \times 10^{-4} \rm cm^{-3}$.
Another WAT radiogalaxy CGCG 128-34 is the brightest member of group 3 (Jaffe \& Gavazzi 1986). 
Mahdavi et al. (2005) report  detection of extended X-ray emission from this group.
Two among the richest groups identified in the H density bin have gas properties similar to small clusters. 
All groups have a velocity dispersion in excess of 200 $\rm km s^{-1}$. \\
Table \ref{tbl-2} gives some information on galaxies in the various density 
regimes. Red and Blue are above or below $g-i = -0.0585 * (M_i +16)+0.78$ \footnote {The cutting line between blue and red has been arbitrarily set 0.2 mag bluer than 
$g-i = -0.0585 * (M_i +16)+0.98$, i.e. the best linear fit to the early-type galaxies.}; 
Lum and Faint are above or below $M_i=-20.5$ mag.

\section{The color-luminosity relation}
\label{colum}
\begin{figure*}[t]
\centering
\includegraphics[width=7.0cm]{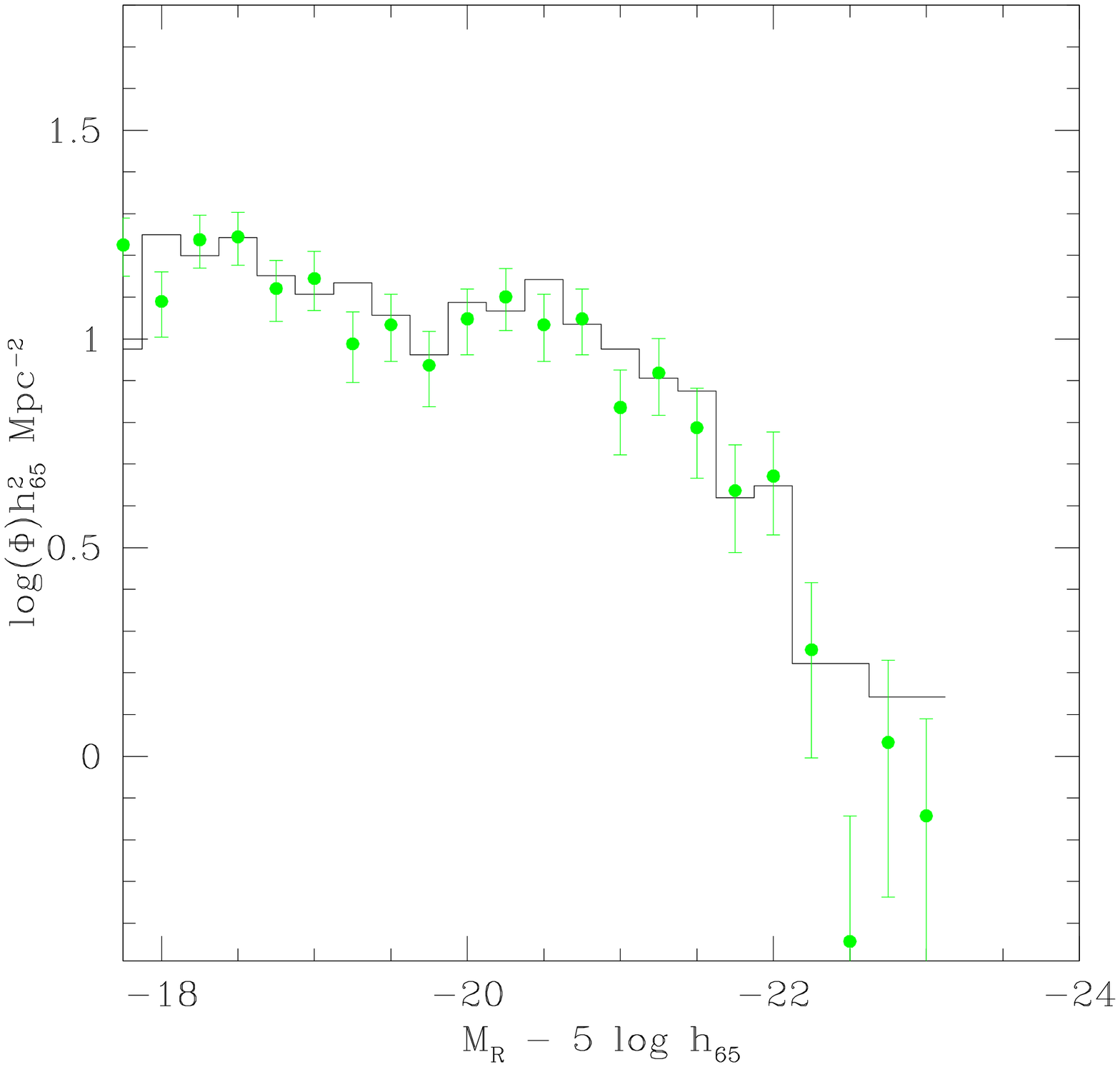}
\includegraphics[width=7.0cm]{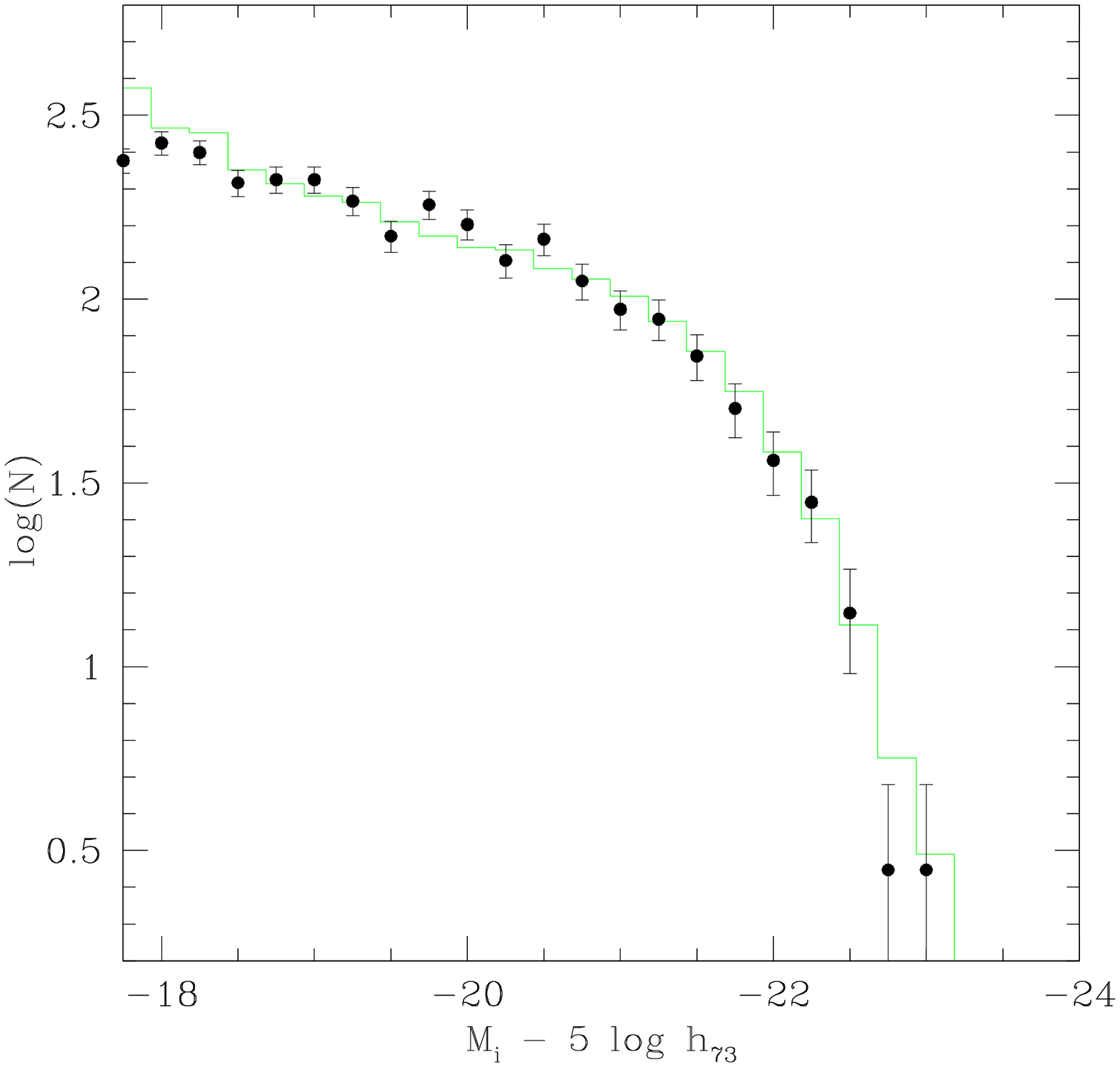}
\caption{(Left panel): the $R$-band luminosity function of the Coma cluster 
adapted from Mobasher et al. (2003) (green dots) and the distribution 
of galaxies in the Coma cluster (including H and UH) from this work (black histogram) 
(converted to R and to $Ho$=65 $\rm km~sec^{-1}~Mpc^{-1}$)
 and normalized according to the areas covered by the two surveys.
 Error bars are omitted to avoid confusion.
(Right panel): the $i$-band luminosity function as derived from Blanton et al. (2005b) for
the general field (green histogram) and the $i$-band luminosity distribution 
in the CS (including L and UL density) from this work (black dots) 
(shifted horizontally to account
for $Ho$=100 $\rm km~sec^{-1}~Mpc^{-1}$ used by Blanton et al. (2005b), and normalized arbitrarily on the vertical axis).
\label{figure4}}
\end{figure*}
The $g-i$ color vs. $i$-band absolute magnitude relation for all galaxies in the 
CS is given in Fig.\ref{figure3}. 
Galaxies are coded according to the Hubble type (top panel), with the late-type 
given alone in the bottom-left panel and the early-type alone in the bottom-right panel.
As expected, the cut in morphological type correlates very well with the separation in color.
In other words the red-sequence, or the locus of points redder than
$g-i = -0.0585 * (M_i +16)+0.78$, is mostly populated by early-type galaxies,
while the large majority of late-type galaxies form a blue cloud mostly at low luminosity,
that steeply becomes redder as the luminosity increases.
 There are however exceptions, i.e. high luminosity spirals often have redder colors than
ellipticals of similar luminosity, as recently emphasized by Lintott et al. (2008) and Bamford et al. (2009). 
These are large, mostly bulge dominated, sometimes
edge-on spirals that are not only reddened by internal absorption, but 
even when their color is corrected for dust absorption (see Fig. 8 of Cortese at al. 2008) they
remain redder then lower luminosity spirals, due to the  prevalence of their intrinsically red bulges, as
concluded by Cortese \& Hughes (2009). 
For example the most luminous spiral galaxy 
near the center of the Coma cluster, the famous 
NGC 4921, in spite of being close to face-on 
(see a plate of this prominent spiral recently taken with the HST by Cook 2006), has $g-i=1.29$, 
i.e. comparable to $g-i$=1.27 and 1.24 of NGC 4874 and NGC 4889 respectively, 
the two central giant ellipticals of Coma. 
Spirals would separate better from ellipticals at high luminosity
if the NUV-H color index was used instead of $g-i$
(as in Gil de Paz et al. 2007; Cortese et al. 2008; Hughes \& Cortese 2009), but 
unfortunately we don't have access to FIR data necessary
to derive the amount of dust obscuration in these galaxies, thus to correct UV magnitudes (see Cortese et al. 2008).

\section{Luminosity distributions}

For the purpose of checking possible incompleteness in our data,
first of all we compute the luminosity distribution obtained from the Sloan $r$-band data
for the Coma cluster alone and we compare it, among the plethora of luminosity functions 
available in the literature, with 
the R-band determination obtained by Mobasher et al. (2003) 
(after converting our $r$(AB) magnitudes into R (Vega) mag and consistently 
adopting their $Ho$=65 km s$^{-1}$ Mpc$^{-1}$). 
Our determination
is normalized to theirs to account for the different portion of the Coma cluster covered by
the two surveys.
The two distributions, given in Fig.\ref{figure4} 
(left panel), agree in details, including
the slight deficit of galaxies near $M_R=-19.5$.\\
Similarly the i-band luminosity distribution of "isolated" galaxies in this work (including 
the UL and L density bins) is compared 
in Fig.\ref{figure4} (right panel) with the luminosity function of the general field 
of Blanton et al. (2005b) (uncorrected for the effect 
of surface brightness, shifted horizontally to account
for the  Blanton et al. (2005b) use of $Ho$=100 km s$^{-1}$ Mpc$^{-1}$, and normalized arbitrarily on the vertical scale).
The two determinations are found in good agreement over the whole luminosity interval, 
supporting the preliminary conclusion that any residual incompleteness of our sample is within few percent.
The slight deficit of galaxies for $M_i>-18.5$ should not influence the conclusions of this work.
In the next section we study the luminosity distribution of galaxies in the Coma supercluster
in bins of local galaxy density and Hubble type.
Due to our inability to correct for possible, even though small, 
residual incompleteness in our data we prefer to give
the number counts in (0.5 mag) bins of $i$-band absolute magnitude (without normalizing them to the sampled volume)
and to call them $luminosity ~distributions$ instead of $luminosity ~functions$. 
The distributions are conservatively plotted for $M_i<-17.75$,
corresponding to $i$ $\leq$17.15 mag, i.e. 0.35 mag brighter than the spectroscopic limit $r$=17.77 
of the SDSS (assuming an average color of $<r-i>=0.25$).\\
We chose the $i$-band instead of $r$-band because it is a better tracer of the stellar mass in galaxies
(Bell et al. 2003).
\begin{figure}[!h]
\includegraphics[width=8.0cm]{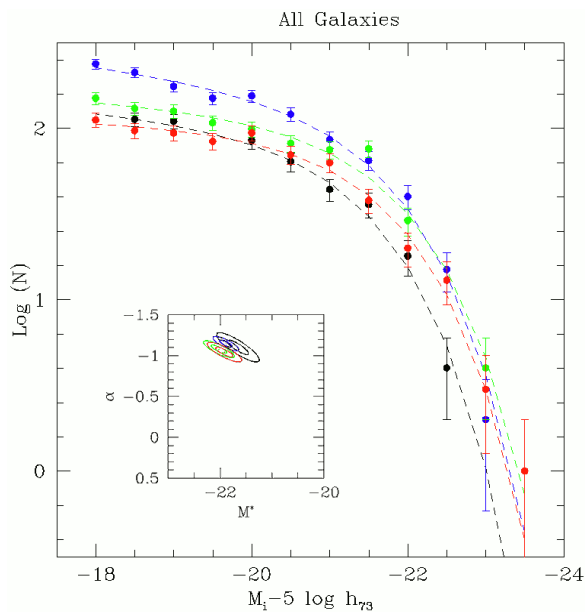}
\caption{$i$-band luminosity distribution of galaxies of all Hubble types coded according to 
bins of $\delta_{1,1000}$ (red=UH, green=H, Blue=L, black=UL)) 
The dashed lines represent Schechter fits to the data whose parameters are given in Table \ref{tbl-3}.
The inset shows the confidence ellipses of the fit parameters. No significant changes are seen in the fit parameters 
with density.
\label{figure5}}
\end{figure}
\begin{figure}[!h]
\includegraphics[width=8.0cm]{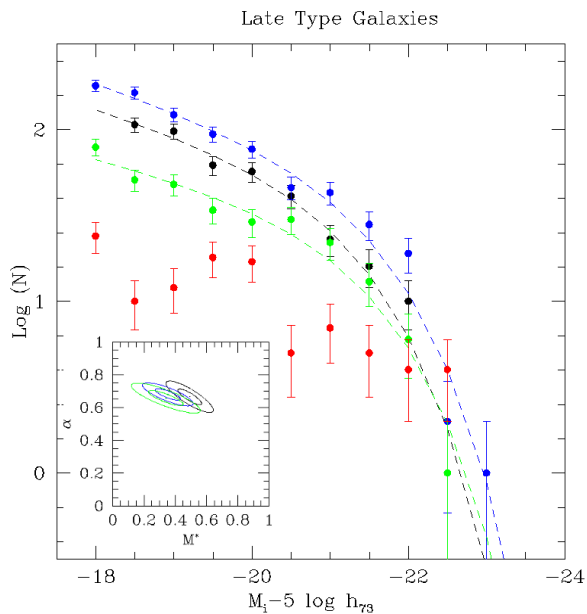}
\caption{Same as Fig. \ref{figure5} for late-type galaxies. No significant changes are seen in the fit parameters 
with density, except for the UH bin where no reliable fit was obtained.}
\label{figure6}
\end{figure}
\begin{figure}[!h]
\includegraphics[width=8.0cm]{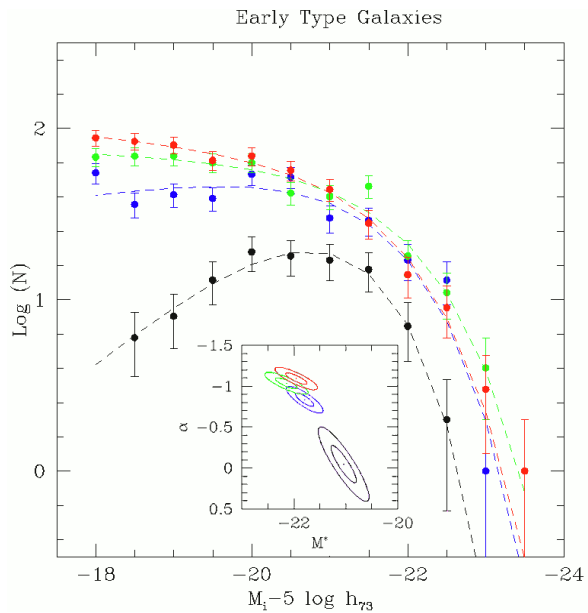}
\caption{
Same as Fig. \ref{figure5} for early-type galaxies. There is a significant decrease of $\alpha$ with increasing density. $M_i^*$ 
stays within one magnitude.}
\label{figure7}
\end{figure}

\subsection{Local density}

The luminosity distributions obtained in four bins of local galaxy density (see Section 3.1) and two of Hubble type are
shown in  Figs. \ref{figure5}, \ref{figure6} and \ref{figure7}.
We have fitted the luminosity distributions obtained in this work
with Schechter functions, computing the best fit parameters with an IDL
code, taking advantage of the MPFIT function, a non
linear least squares fitting task based on the Levenberg-Markwardt 
algorithm (Mor\'e 1978; Markwardt 2009). We have also obtained the covariance
matrix of the parameters again from the MPFIT function and we have
evaluated the confidence ellipses at 1 and 2 $\sigma$ (as shown in the insets of 
Figs. \ref{figure5}, \ref{figure6} and \ref{figure7}).
The fit parameters are given in Table \ref{tbl-3}.\\
Fig.\ref{figure5} shows that the distributions obtained mixing together galaxies of all Hubble 
types are completely insensitive to the environment.
From Table \ref{tbl-3} it is apparent that $\alpha$ and $M_i^*$ are consistent within the errors
in all density bins. This is very much the consequence of a cosmic conspiracy, since 
the distributions are complementary with respect to the Hubble type, 
as we will show by separating late-type galaxies from early-types 
(one can almost obtain Fig. \ref{figure7} by reversing the symbol-colors in Fig. \ref{figure6})
\\
When we consider the late-type galaxy population alone (Fig.\ref{figure6}) there is an evident
decrease in the number of objects with increasing $\delta_{1,1000}$. 
In the highest density bin (UH) the number of late-type galaxies is so small that it is
even insufficient for fitting a Schechter function. 
 \begin{table}[!t]
  \caption{Parameters of the Schechter functions
  fitted to the luminosity distributions.\label{tbl-3}}
  \begin{tabular}{llccc}
  \hline
  \hline
      &  & $\Phi$ & $\alpha$ & $M_i^*$ \\ 
   \hline
  all        & UH &  103 $\pm$ 16&    -1.04 $\pm$ 0.05&    -21.91 $\pm$ 0.16\\
             &  H &  120 $\pm$ 17&    -1.07 $\pm$ 0.05&    -22.02 $\pm$ 0.14\\
             &  L &  156 $\pm$ 20&    -1.14 $\pm$ 0.04&    -21.89 $\pm$ 0.12\\
             & UL &  100 $\pm$ 21&    -1.10 $\pm$ 0.08&    -21.65 $\pm$ 0.20\\
  \hline
   early    & UH &    72 $\pm$ 14&    -1.09 $\pm$ 0.06&    -21.95 $\pm$ 0.20\\
       	    &  H &    69 $\pm$ 13&    -1.03 $\pm$ 0.06&    -22.14 $\pm$ 0.20\\
            &  L &    79 $\pm$ 13&    -0.83 $\pm$ 0.08&    -21.79 $\pm$ 0.17\\
            & UL &    56 $\pm$	7&    -0.04 $\pm$ 0.22&    -21.03 $\pm$ 0.24\\
  \hline
   late     & UH &        -       &       -  &       - \\
            &  H &    29 $\pm$  10&   -1.28 $\pm$ 0.10&    -21.80 $\pm$ 0.30\\
            &  L &    59 $\pm$  13&   -1.37 $\pm$ 0.06&    -21.84 $\pm$ 0.19\\
            & UL &    52 $\pm$  17&   -1.33 $\pm$ 0.10&    -21.58 $\pm$	0.27\\
  \hline
  \end{tabular}
  \\
  \end{table}
The other two fit parameters $\alpha$ and $M_i^*$ are
consistent within the errors (in the three lowest density bins we obtain a faint-end slope
$\alpha \sim 1.3$ and $M_i^* \sim -21.7$), implying that the shape of the luminosity distribution 
of the late-type galaxy population is 
relatively insensitive to the local galaxy density, except for the highest density environments where
late-type galaxies are rare at any luminosity.
The luminosity distribution of early-type galaxies is given in Fig. \ref{figure7} divided in the usual bins
of local density.
In the three highest density bins (L, H, UH) the luminosity distributions display small but continuous changes:
their Schechter fit parameter $\alpha$ increases marginally and $M_i^*$ stays constant within the errors (see Table \ref{tbl-3}).
Highly significant discrepant $\alpha$ is found in the lowest density bin (UL or $\delta_{1,1000}<0$), 
but the luminosity distributions of early-type galaxies have negative  faint-end slopes in the two lowest density bins. 
This reflects a genuine lack of low luminosity early-type 
\begin{onecolumn}
      \begin{figure*}[!t]
      \centering
      \includegraphics[width=7.9cm]{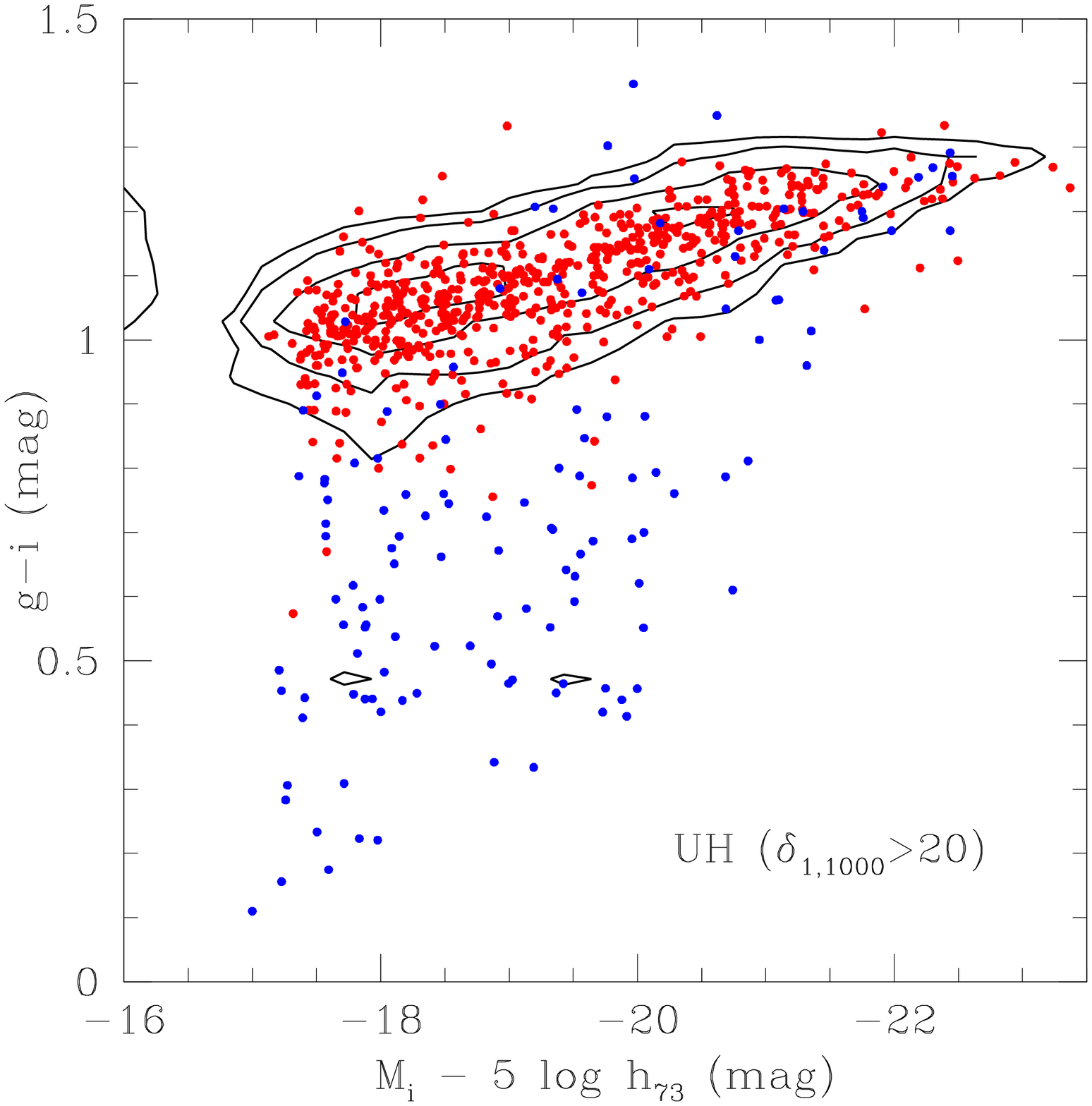}
      \includegraphics[width=7.9cm]{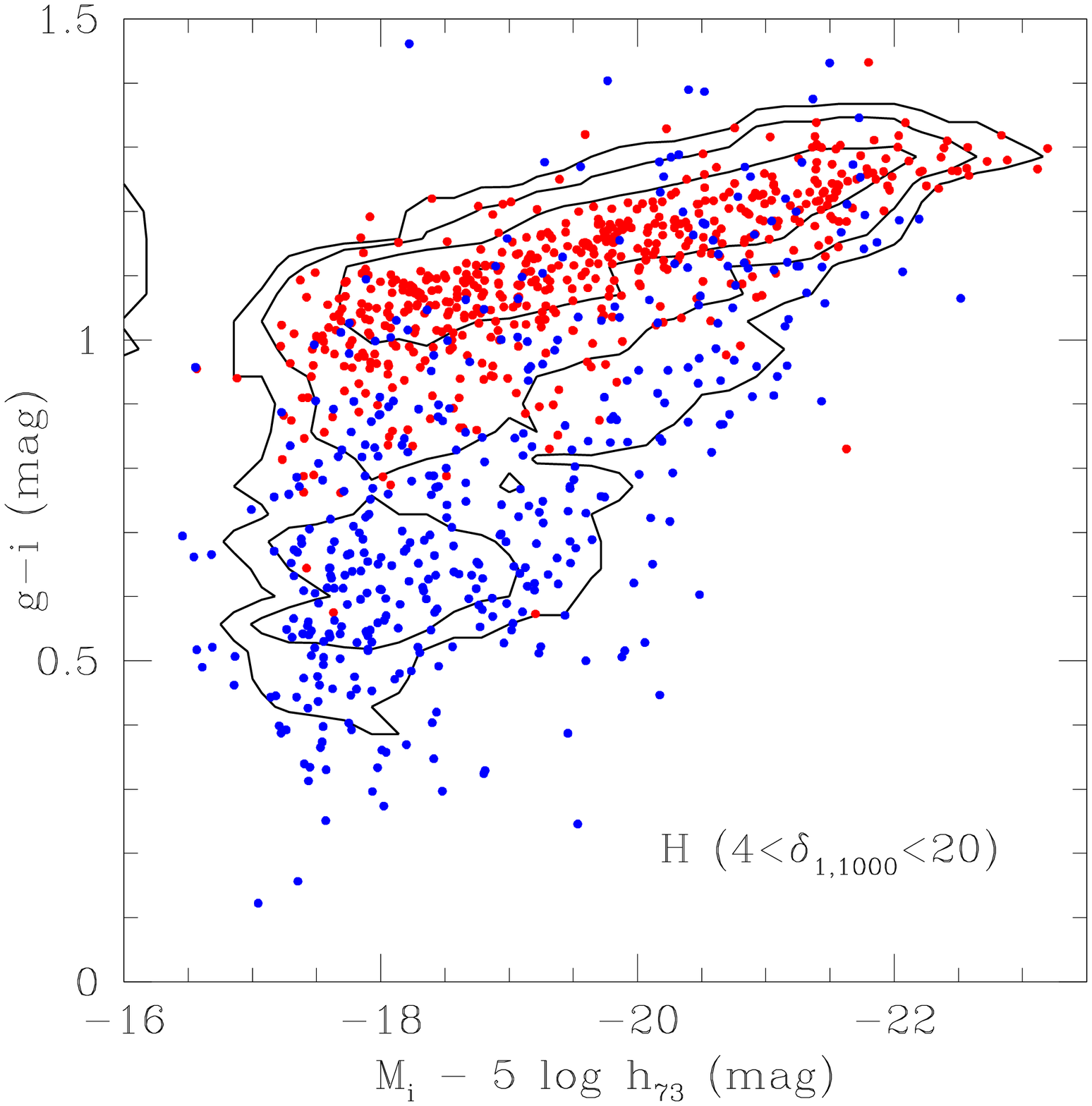}\\
      \includegraphics[width=7.9cm]{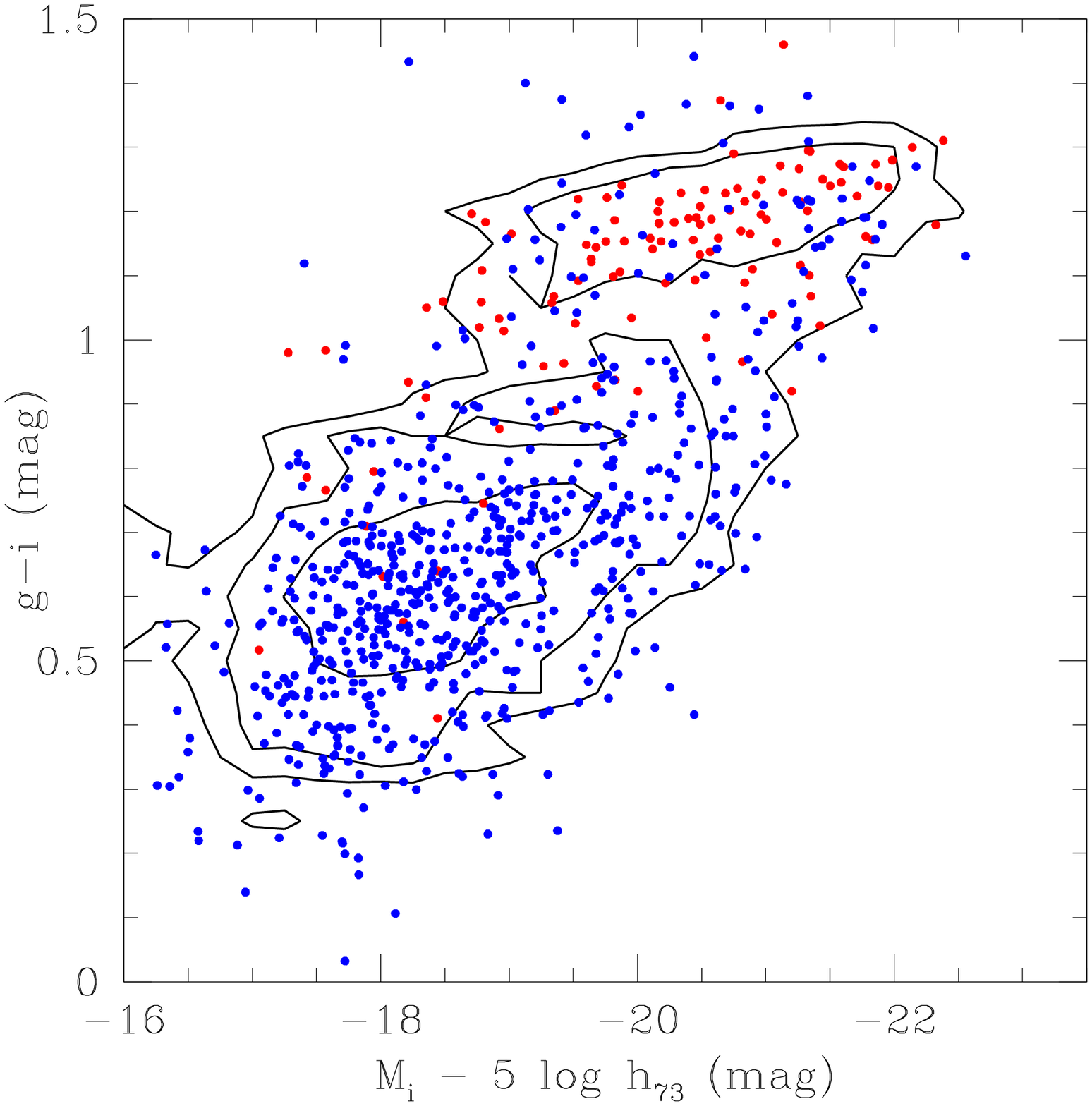}
      \includegraphics[width=7.9cm]{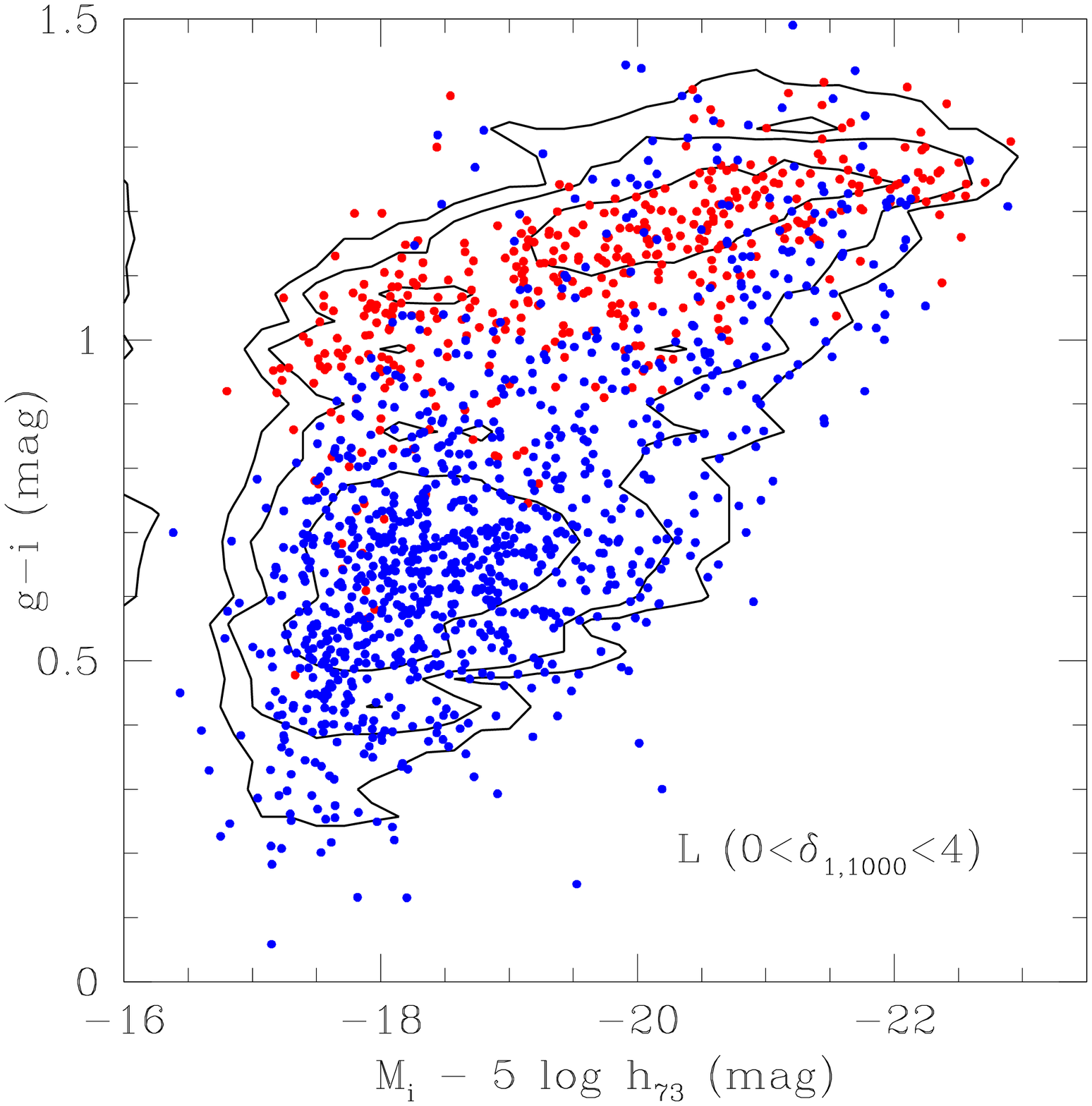}
      \caption{$g-i$ color versus $i$-band absolute magnitude relation for galaxies in the CS,
      from UH to UL clockwise from top left, coded according 
      to galaxy type (blue = late, red= early). Contours of equal density are given. 
      \label{figure8}}
      \end{figure*}
\begin{table}
\centering
    \caption{Parameters of the Gaussian fitted to the $\Delta(g-i)$ histograms in Fig. 
    \ref{fig11b} for $M_i>-20.5$.}
  \begin{tabular}{lcccc}
  \hline
  \hline
      		\label{fig13}   & Late   &                             & Early   &  \\
        Dens    & $<\Delta(g-i)_{L}> $     & $\sigma_{L}$ & $<\Delta(g-i)_{E}> $ & $\sigma_{E}$\\
   \hline
UH &           -          -     &	-     &    -0.004 $\pm$   0.001 &      0.049\\
H &       -0.365 $\pm$    0.003 &	0.193 &     0.006 $\pm$   0.001 &      0.056\\
L &       -0.409 $\pm$    0.001 &	0.162 &    -0.004 $\pm$   0.001 &      0.073\\
UL &      -0.444 $\pm$    0.001 &	0.136 &    -0.001 $\pm$   0.003 &      0.063\\
  \hline
  \label{tab4}
  \end{tabular}
  \\
  \end{table}      
\end{onecolumn} 
\begin{twocolumn}
galaxies (dEs) in the low density cosmic web (UL). 
At high luminosity instead, even in the lowest density regime it appears 
that the giant elliptical galaxies are well formed, although $M_i^*$ at UL is probably 0.5-1 magnitude fainter than 
at larger densities.\\
Summarizing, Figs. \ref{figure5}, \ref{figure6} and \ref{figure7} show that the bright end of both the late- and the early-type sequences
are consistently formed in all environments (small differences in $M^*$ are due to the 
small number statistics in the brightest bins). 
On the opposite the fraction of faint early-type galaxies (dEs) depends on the environment, 
being significantly higher in dense than in isolated regions. 
The one of late-types does the opposite, in agreement with
Balogh et al. (2004a), 
Baldry et al. (2006), Mart{\'{\i}}nez \& Muriel (2006), Haines et al. (2006, 2007).
Among isolated galaxies of low luminosity the missing early types are compensated by the abundant late-types. 
In dense regions the relative population mix is reversed.\\
It is remarkable that the luminosity distributions are consistent at the high luminosity end
in all environments, suggesting that 
neither early-type (e.g. Whiley et al. 2008) nor late-type (e.g. Boselli et al. 2006) 
massive galaxies are currently undergoing major
transformations. These have probably been going on at significantly
higher redshift (Faber et al. 2007).

\section{The environmental dependence of the color-luminosity relation}
\label{coldensity}

Among the most relevant achievements of the SDSS there is the discovery that
the color (e.g. Blanton et al. 2005c, Haines et al. 2007) and the 
specific star formation rate (e.g. Kauffmann et al. 2004) of galaxies are strongly influenced by
the environment. 
\begin{figure}
\centering
\includegraphics[width=8cm]{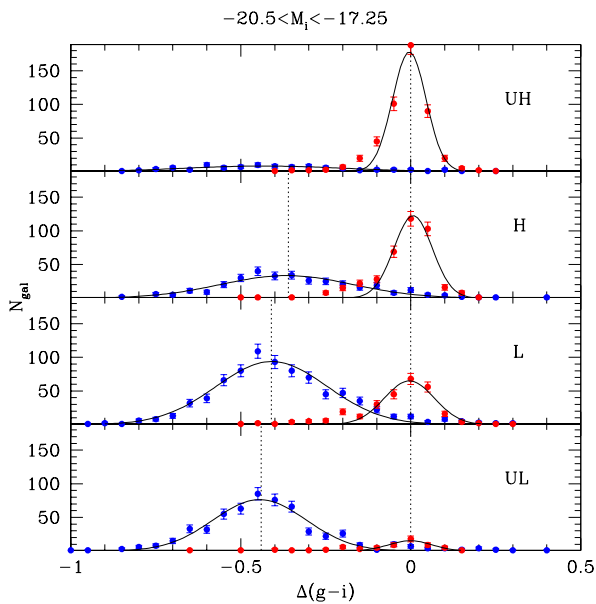}
\caption{Histograms and Gaussian fits of $\Delta(g-i)$, representing the difference in color of individual galaxies 
from the linear fit to the red sequence, for galaxies in four density bins and $M_i>-20.5$.
Late-type (blue symbols) are sparated from early-type galaxies (red symbols). 
\label{fig11b}}
\end{figure}
The Coma supercluster offers a very direct snapshot of these environmental transformations, as illustrated  
in Fig. \ref{figure8}, showing the 2-Dimensional color-luminosity relations derived
in four bins of local galaxy density (UL, L, H, UH).
The lowest density bin
is composed almost entirely (see Table \ref{tbl-2}) of blue (78\%), late-type (82\%), dwarfish galaxies, 
however even in this bin 
a hint of red sequence exists (22\%) composed of early-type galaxies (18\%), but is significantly lacking at low-luminosity
($M_i>-18.5$).
As the local density approaches $0<\delta_{1,1000}<4$, i.e. the regime traced by small groups and other 
faint structures in the Great Wall, the blue sequence remains dominating (61\%), the red sequence appears 
well formed in its full luminosity span, but early-type galaxies represent only 35\% of all galaxies.
By increasing the galaxy density to $4<\delta_{1,1000}<20$, dominated by large groups and by the cluster outskirts,
the percentage of late type galaxies drops to 39\% and that of early-type increases to 61\%. 
At the highest density $\delta_{1,1000}>20$, i.e. in the core of the rich clusters, 
galaxies consist of nothing but early-type objects (84\%) distributed along the red sequence (82\%). 
We wish to emphasize that some red sequence galaxies exist in all environments but they span the full luminosity 
range $-17.75<M_i<-23$ (including dEs) only for densities larger than $\delta_{1,1000}>0$ and that
the mix of morphological type depends on the local galaxy density
(see also Fig. \ref{fig13}).\\
Figure \ref{fig11b} shows histograms of the $\Delta(g-i)$ color, e.g. the difference in color of individual galaxies 
from the linear fit to the red sequence (see note 3), in other words after removing the slope of the red-sequence from
the color-luminosity relation of Fig. \ref{figure3},
in four bins of galaxy density and for $M_i>-20.5$ mag (where the majority of the late-type galaxies are found). 
The distributions of early and late-type galaxies are fitted
with two Gaussians, whose parameters are listed in Table \ref{tab4}. 
The ratio of late to early objects is strongly density dependent.
Moreover the mean color difference of the late-type population with respect to the early-type sequence
reddens systematically from UL ($\Delta (g-i)$ =-0.44 mag),
L ($\Delta (g-i)$ =-0.41 mag), H ($\Delta (g-i)$ =-0.36 mag) (the contribution of the blue sequence at 
UH is so marginal that no Gaussian fitting is derived),  indicating that galaxies, 
even though remaining in the blue sequence, migrate to redder colors, i.e. 
are subject to progressive quenching of the star formation as the environment density progressively increases.

\section{Discussion}

\begin{figure*}
\centering
\includegraphics[width=13.0cm]{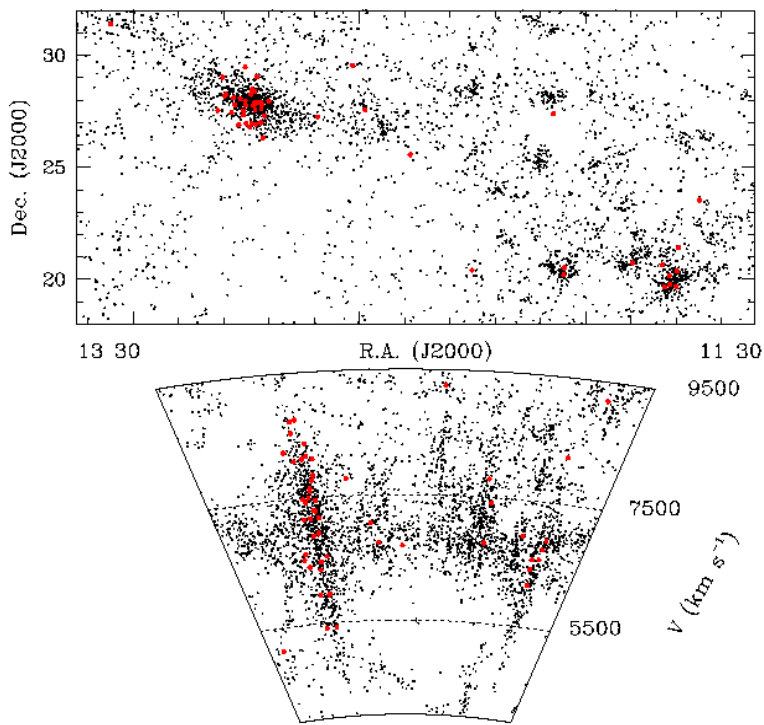}
\caption{The projected distribution of the 53 PSB galaxies selected according to eq. \ref{eqPSB} (large red dots) in 
the Coma Supercluster (small dots). 
\label{fig11}}
\end{figure*}

At the present cosmological epoch there is evidence that dynamical units (clusters) are
continuously fed by the infall of clouds predominantly constituted by late-type galaxies occurring along the filamentary structures
that compose the cosmic web (Adami et al. 2005, Cortese et al. 2008; Tran et al. 2008),
as predicted by simulations of the hierarchical galaxy formation paradigm (e.g Springel et al. 2005).
Direct evidence for infall into the Coma cluster derives from X-ray observations supporting the recent
infall of the NGC4839 group (Neumann et al. 2003) onto the main cluster. 
More indirect line of evidence for infall of late-type galaxies on the Coma cluster is in our own data, 
i.e. the velocity distribution of late-type galaxies belonging to the Coma cluster (in the density bin UH and H, corresponding
to the core + outskirts) is significantly non Gaussian
with $\sigma_V$ = 1046 $\rm km s^{-1}$, exceeding $\sigma_V$ = 872 $\rm km s^{-1}$ of early-type members
which, instead have isotropic distribution.\\ 
The way the color-luminosity plane is populated by $\sim$ 4000 galaxies representative
of different environments in the local universe around the Coma cluster, 
as revealed by the Sloan survey, strengthens the following scenario: 
\begin{enumerate}
\item No strong environmental differences are seen in the galaxy population of the local universe (Fig. \ref{figure5}) unless 
late-type galaxies (Fig. \ref{figure6}) are counted separately from early-types (Fig. \ref{figure7}):
the former decrease in fraction with increasing local density, viceversa the latter. \\
\item  The differences at $z$=0 involve mostly galaxies of low luminosity (mass) (well below $M^*$), 
in agreement with Scarlata et al. (2007). The missing low-luminosity late-type galaxies in dense environments
compensate exactly the missing low-luminosity early-type galaxies in loose environments.\\
\item  High luminosity galaxies ($M_i<-20.5$) show little environmental differences in both
their early-type and late-type components (see Table \ref{tbl-3}), i.e.
early-type galaxies have $M^*$ $\sim$ 1 mag fainter in extreme isolation. 
This shift shows up even in the global Luminosity Function (all types) as a $\sim$ 0.5 mag difference in $M^*$
of UL compared to higher density. 
A similar shift is found in the zCOSMOS low redshift bin ($0.2<z<0.5$) by Tasca et al. (2009) 
who refers to Bolzonella et al. (2009). Notice however that their study covers a density regime
that falls entirely within our UL and L bins and misses high density Coma-like clusters.
\end{enumerate}
Points 1-3 form a coherent evolutionary picture assuming that 
at recent cosmological epochs strong environmental transformations ("Nurture") of the galaxy 
population involve low luminosity blue amorphous dwarf galaxies (BCDs) 
that progressively become dEs in their infall into denser structures, in further agreement with Haines et al. (2008).
This major shift consists of a pure color migration from blue to red by
about 0.5 mag (g-i), causing a major spectroscopical change, accompanied by a minor change in morphology.
No significant luminosity growth is seen along this process, as derived from the comparable luminosity
of blue systems abundant in isolation and of red dwarfs abundant in denser regions.\\
High luminosity galaxies appear to be already in place, irrespective of the environment, from previous cosmological epochs,
probably as a consequence of the "Nature" process known as downsizing   
(Cowie et al. 1996; Gavazzi et al. 1996; Gavazzi \& Scodeggio 1996; Boselli et al. 2001, Fontanot et al. 2009).
Massive ellipticals and massive bulge-dominated spirals are likely to have formed at significant redshift due to major merging 
(as predicted by merger trees, e.g. De Lucia et al. 2006 and observed by Faber et al. 2007, Scarlata et al. 2007, Bell et al. 2007).\\
At progressively lower redshift ($z<1$) no evolution is observed for the massive early-type galaxies 
(e.g. Whiley et al. 2008; Cool et al. 2008) and 
the formation of the red sequence is shifted to lower luminosities
in the environment of over-dense regions (groups and clusters).\\
The evolutionary scenario emerging from the present investigation of galaxies at $z$=0 
is coherent with the results of De Lucia et al. (2007) who investigated several clusters at high redshift ($0.4<z<0.8$).
They find that the ratio of luminous to faint galaxies in the red sequence increases with $z$, and conclude that 
the low luminosity part of the red sequence was formed only between $z$=0.5 and $z$=0
\footnote{By adjusting arbitrarily the limit between luminous and faint galaxies in our survey to $M_i=-20.5$
we obtain that the ratio lum/faint in the red sequence is 0.32 in the UH bin (dominated by the Coma cluster),
deliberately matching the value found for the Coma cluster by De Lucia et al. (2007) using V-band observations.
This ratio is strongly environment dependent at $z$=0 as it increases from 0.32 in UH, 0.47 in H,
0.56 in L to 0.82 in UL (see Table 2). This growth with decreasing density mimics remarkably the growth with increasing 
$z$ found by De Lucia et al. (2007)}.\\
At $z$=0 our analysis seems to indicate that major merging is not necessarily the relevant environmental process for the
build-up of the red sequence at low luminosity.
Instead we concur with previous investigations that low mass amorphous blue objects are transformed into dEs by
migration across the green valley, due to quenching of the star formation, primarily
in galaxies falling into the clusters and to a lesser extent in groups and looser structures.
The physical mechanisms that appear most likely to be behind the migration
are harassment (Moore et al. 1999), ram-pressure stripping (Gunn and Gott, 1972), 
as proposed by Boselli et al. (2008a,b), confirmed by Hughes \& Cortese (2009) for the Virgo cluster,
and recently extended to the Perseus cluster by Penny et al. (2010).
According to the latest simulations of ram-pressure by Bekki (2009), 
gas stripping with consequent gradual truncation of the star formation occurs
even in small groups, but it proceeds less efficiently than in cluster environments, where
low-mass galaxies are likely to truncate their star formation more rapidly. 

\subsection{Post-Star-Burst Galaxies}
\label{PSB}

In agreement with Boselli et al. (2008a)
we propose that Post-Star-Burst (PSB) galaxies represent the current tracer of the above migration.
k+a (also named E+A) galaxies, identified by their strong Balmer absorption lines, 
gained attention after their discovery in distant clusters by Dressler \& Gunn (1983)
and in the Coma cluster by Caldwell et al. (1993).
They were interpreted as Post-Star-Burst galaxies in the seminal work by Couch \& Sharples (1987).
Nowadays they are known to have rapidly ceased to form stars between $5\times 10^7$ and $1.5\times 10^9$
years prior to observations (see Poggianti et al. 2009 and references therein). 
Significant downsizing is observed in these systems as their typical
luminosity is $M_V \leq -20$ in clusters at $z$=0.5, compared with $M_V \geq -18.5$ 
found in the Coma cluster (Poggianti et al. 2004). 
The youngest (Blue) PSB galaxies are found
to correlate with the X-ray structure of the Coma cluster, suggesting that the star formation
has been suppressed as a consequence of the hostile environment for those galaxies that have
recently accreted (Dressler et al. 1999; Poggianti et al. 1999). 
Studies at $0.4<z<0.8$ confirm that
k+a galaxies are preferentially found in dense environments (Poggianti et al. 2009), however
this point is controversial (e.g. Zabludoff et al. 1996, Balogh et al. 2005).\\
We find 53 such galaxies in our survey (meeting the criterion for k+a given in Section 2). 
Their distribution 
appears significantly clustered. In fact we find only 2 PSBs in the UL bin (0.5\%), 6 in the L bin (0.8\%), 
13 in the H bin (2.4\%), predominantly in the outskirt of the Coma cluster, and
the remaining 32 belonging to the densest  UH bin (7.3\%), as illustrated in Fig.\ref{fig11} and in Table \ref{tbl-2}
\footnote{The fraction of PSB galaxies in Table \ref{tbl-2} has been computed dividing the number of PSBs by
 the number of galaxies with signal-to-noise ratio $>$5 in the spectra.}.
The color of PSB galaxies tend to fill the gap between the red and the blue sequence (green valley), as shown in Fig.\ref{figure11}.
In agreement with Poggianti et al. (2004) their distribution is skewed toward low luminosity, 
(43/53 PSBs have $M_i> -19.5$). 
\begin{figure}
\includegraphics[width=8.0cm]{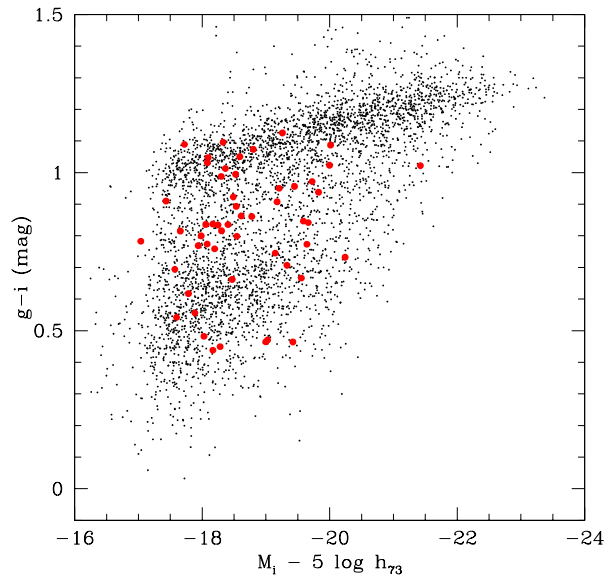}
\caption{$g-i$ color versus $i$-band absolute magnitude relation for all SDSS galaxies 
(small black dots) and for 90 CGCG galaxies without spectral classification (large black dots).
Red dots represent PSB galaxies selected according to eq.\ref{eqPSB}.
\label{figure11}}
\end{figure}
 \begin{figure}
 \includegraphics[width=8.0cm]{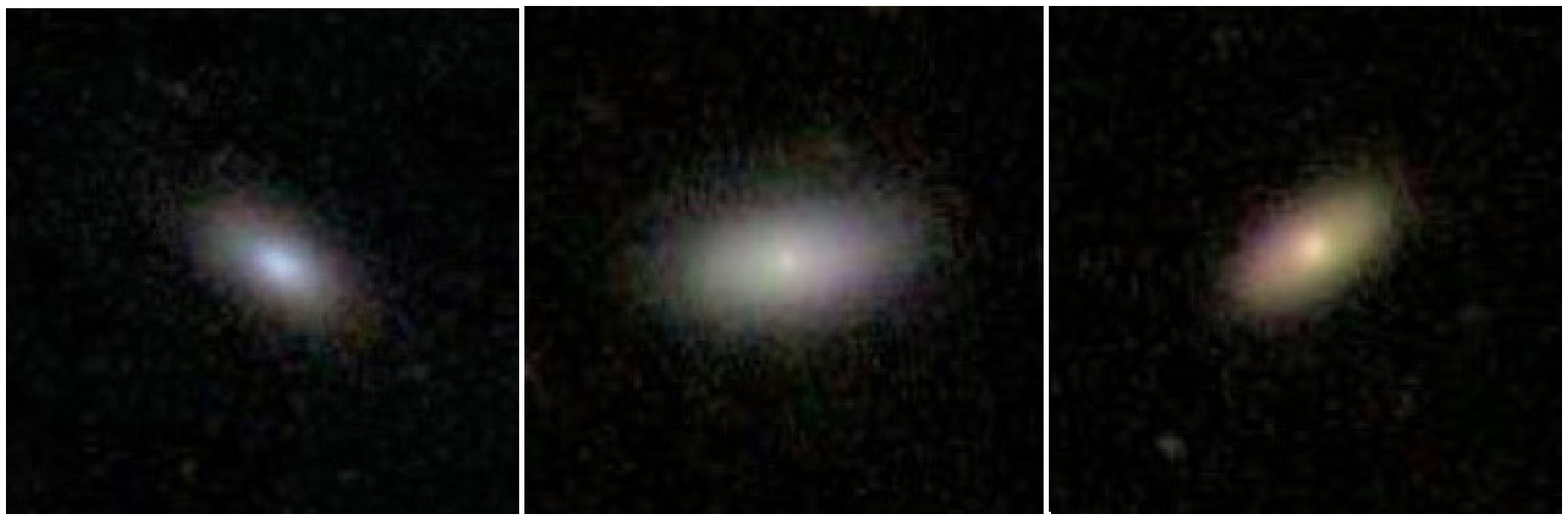}
 \includegraphics[width=8.0cm]{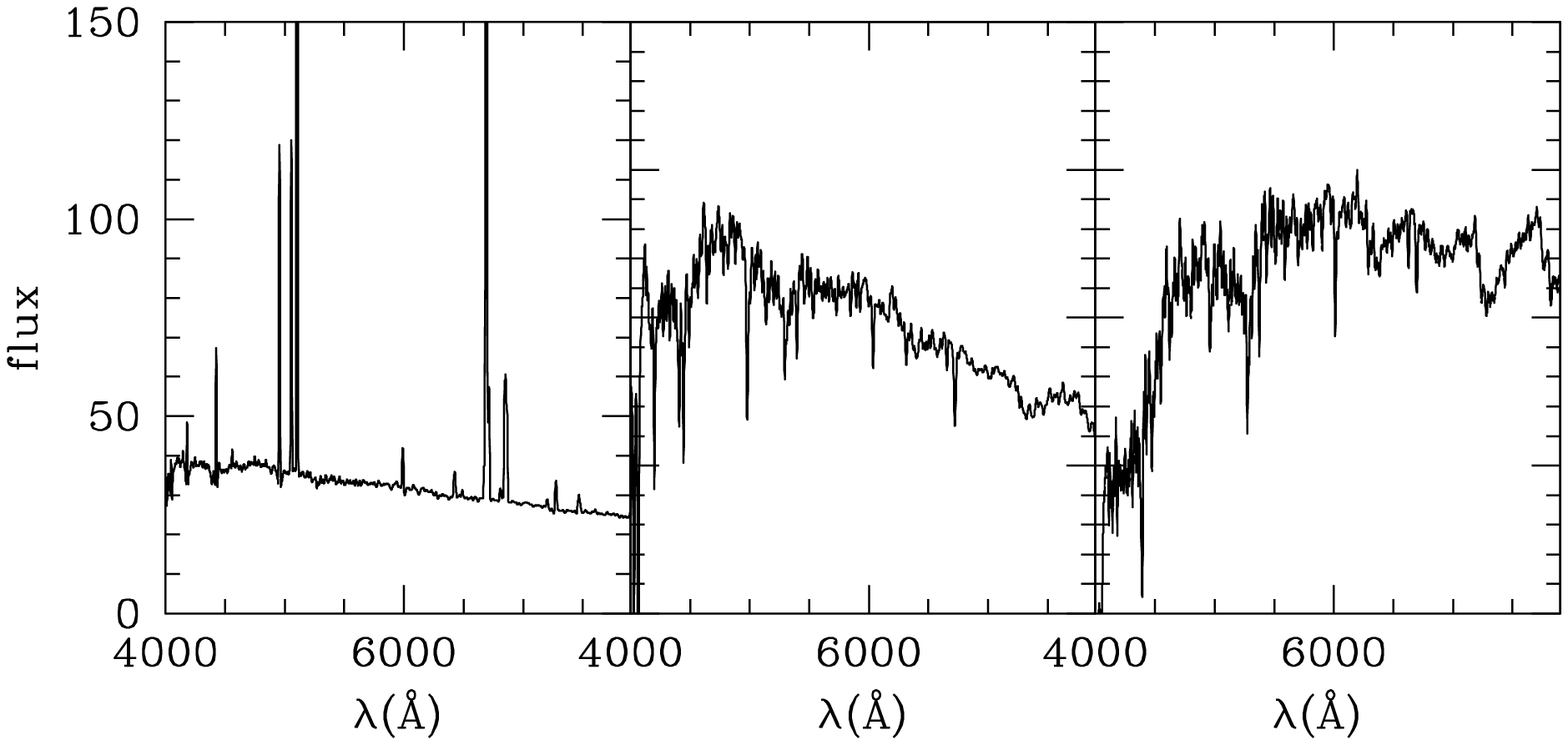}
 \caption{images and spectra of (left to right) one BCD, one PSB with blue continuum and one dE galaxy from this work.
 \label{fig12}}
 \end{figure}
We can exclude that the spectral incompleteness at high luminosity 
discussed in Section \ref{spec} biases against luminous PSBs 
since among luminous galaxies ($M_i< -20.5$) we expect to miss only 0.1$\pm$0.1 PSB.\\
PSB galaxies are likely to represent the transition systems between dwarf star forming and quiescent galaxies. 
They have similarly amorphous  morphology, as shown in Fig.\ref{fig12},
where three galaxies with similarly low luminosity ($-18.3>M_i> -19.7$) are presented, 
one randomly selected among the plethora of
isolated BCDs, one among the blue PSBs and one of the dEs in the Coma cluster. 
In a black-white picture they would result absolutely indistinguishable. 
Their distinctive character, represented by the color of their continuum 
and by the intensity of their lines, resides in their history of star formation, that is currently
active in BCDs, has ceased less than 1.5 Gyr ago in PSBs (0.5 Gyr in the blue ones) and is absent among dEs.\\
Figure \ref{fig13} gives the fraction of PSBs and of early-type galaxies as a function of 
the log of the local density (left panels) and of the projected distance from the Coma cluster (right panels). 
The fraction of PSBs stays constant (approximately at the 1\% level) up to $\delta_{1,1000}$ $\sim$ 10, 
and reaches
8\% at $\delta_{1,1000}$ $\sim$ 30, without further increasing in the cluster core. The radial distribution
around the Coma cluster shows a maximum at $\sim$ 1-2 Mpc projected radius and does not increase further inward.
At this projected distance from the center, the density of the hot gas estimated by Makino (1994) is 
n $\sim 7\times 10^{-4} \rm cm^{-3}$, sufficient to produce significant stripping according to Bekki (2009).

\section{Conclusions and summary}

If our assumption that PSBs are the carriers of the transformation occurring across the green valley holds true, 
their strong clustering around the densest structures is in apparent contradiction with the fact that
the migration discussed in Section \ref{coldensity} involves also much less dense environments.
Timescales offer an obvious solution to this riddle.
To see the PSB signature in a galaxy spectrum, it is required that the time-scale of the truncation of 
the star formation should be
as short as $\sim$100 Myr, as concluded by Boselli et al. (2008a) who model a highly efficient ram-pressure
event in the Virgo cluster. If gas ablation occurs with longer timescales, the star formation results gradually
suppressed, with resultant persistence of residual Balmer lines in emission, thus without PSB signature.
Ram pressure stripping in clusters is a fast phenomenon, while the same mechanism in groups (Bekki 2009) and  
other related mechanisms, such as harassment (Moore et al. 1999) 
or starvation occur in looser aggregates with longer time scales. \\
The fact that clusters are fully evolved since $z$=1 makes them the most favorable environments for the
transformations at recent epochs. At larger look back time the action was going on at lower rates in less dense
environments, inseminating the red sequence with small amounts of late-type galaxies (see Figure \ref{figure8},
bottom panels), but since clusters 
have fully formed, the Nurture effect became more vigorous in their environment, leading to an almost
complete draining of blue galaxies (see Figure \ref{figure8}, top panels).
\begin{figure*}[!t]
 \centering
\includegraphics[width=8.cm]{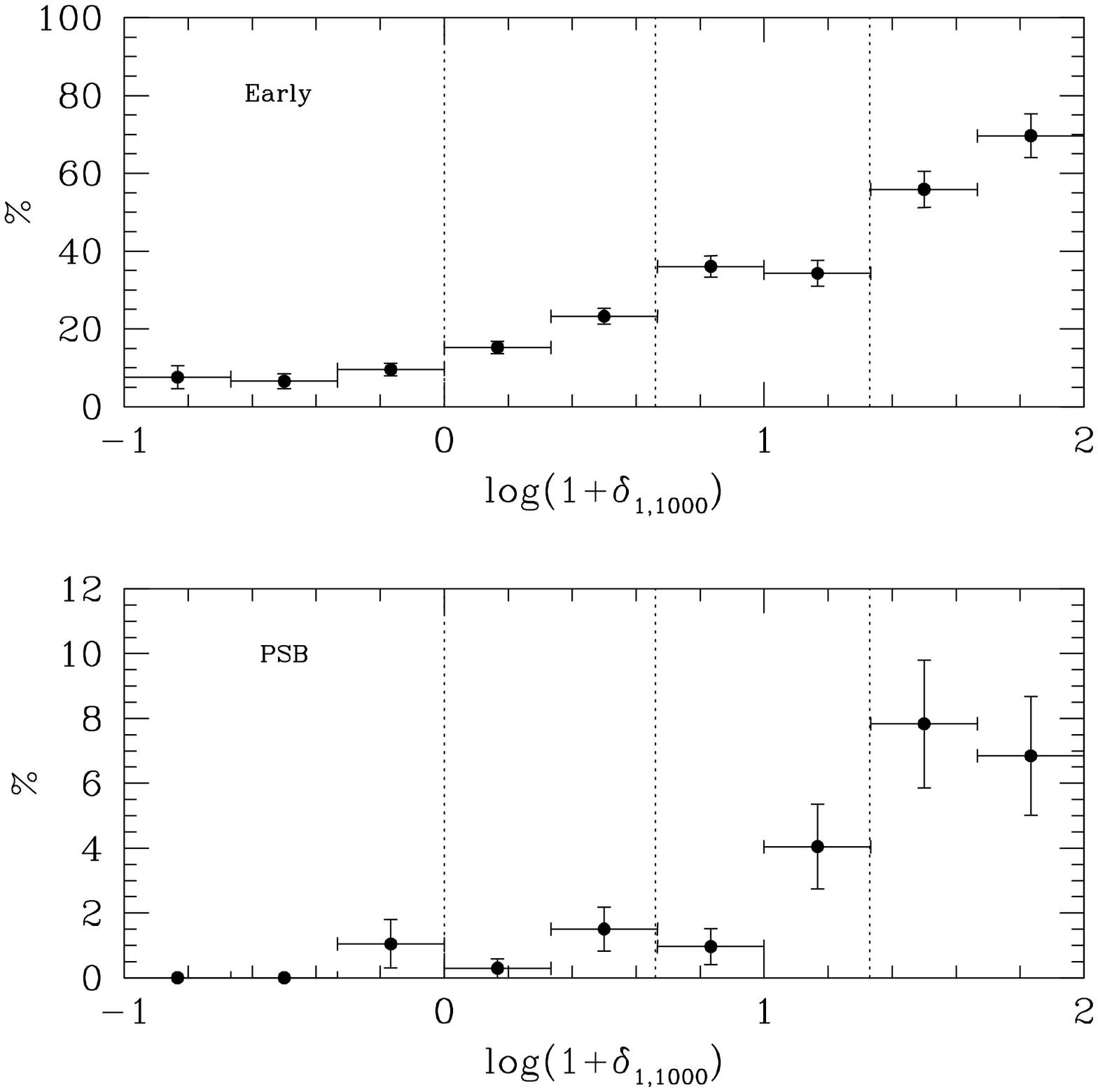}
\includegraphics[width=8.cm]{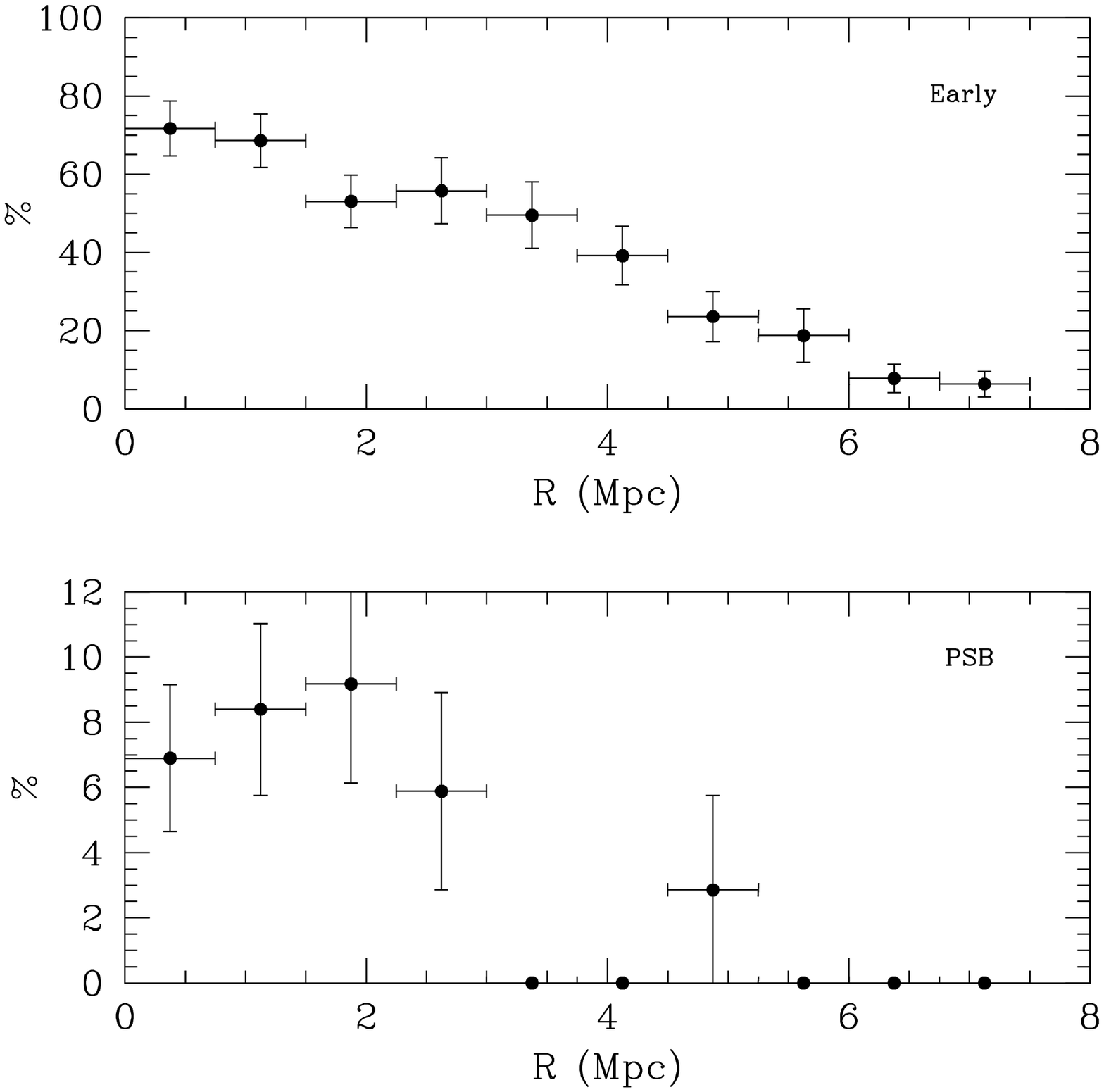}
\caption{Fraction of of early-type galaxies (top) and of PSB (bottom) 
as a function of the (logarithmic) $\delta_{1,1000}$ (left) and of 
the radial distance from the Coma cluster (right). 
The dashed vertical lines mark the density thresholds adopted in this work.
\label{fig13}}
\end{figure*}

The main results of the present analysis of $\sim$ 4000 galaxies in the Coma Supercluster, carried out using
DR7-SDSS data can be summarized as follows:
\begin{enumerate}
\item At $z$=0 there is evidence of little luminosity evolution among high luminosity galaxies ($M_i< -20$).
\item At $z$=0 there is strong migration of late/blue/star forming into early/red/quiescent systems
taking place among dwarf ($M_i> -19$) galaxies, as summarized in Fig. \ref{fig14}.
\item Although the migration takes place also in loose structures it is more efficient in denser
environments, such as clusters of galaxies ("Nurture").
\item Quenching of the star formation is induced in clusters by fast gas-ablation mechanism,
namely ram-pressure.
\item Post-Star-Burst galaxies appear the tracers of the migration taking place in dense environments. 
\end{enumerate}
\begin{figure}
\centering
\includegraphics[width=8.0cm]{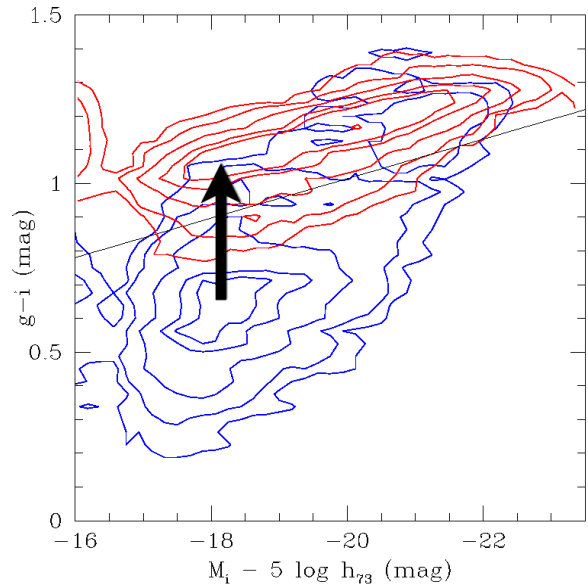}
\caption{The $g-i$ color versus $i$-band absolute magnitude relation for late-type galaxies (blue contours)
and for early-type galaxies (red contours). The arrow marks the effect of Nurture at $z$=0 in dense environments. 
Low-luminosity blue galaxies have their star formation truncated as a result of gas stripping.
\label{fig14}}
\end{figure}

\begin{acknowledgements}
We thank Roberto Decarli for help in querying the SDSS archive and Veronique Buat, Gabriella De Lucia, 
Michele Fumagalli, Bianca Poggianti and Marco Scodeggio for helpful discussions.
We are grateful to Paolo Franzetti and Alessandro Donati for their contribution to GOLDMine, 
the Galaxy On Line Database extensively used in this work (http://goldmine.mib.infn.it).
We acknowledge the constructive criticism from an unknown referee.
The present study could not be conceived without the DR7 of SDSS. 
Funding for the Sloan Digital Sky Survey (SDSS) and SDSS-II has been provided by the 
 Alfred P. Sloan Foundation, the Participating Institutions, the National Science Foundation, 
 the U.S. Department of Energy, the National Aeronautics and Space Administration, 
 the Japanese Monbukagakusho, and 
 the Max Planck Society, and the Higher Education Funding Council for England. 
 The SDSS Web site is http://www.sdss.org/.
 The SDSS is managed by the Astrophysical Research Consortium (ARC) for the Participating Institutions. 
 The Participating Institutions are the American Museum of Natural History, Astrophysical Institute Potsdam, 
 University of Basel, University of Cambridge, Case Western Reserve University, The University of Chicago, 
 Drexel University, Fermilab, the Institute for Advanced Study, the Japan Participation Group, 
 The Johns Hopkins University, the Joint Institute for Nuclear Astrophysics, the Kavli Institute for 
 Particle Astrophysics and Cosmology, the Korean Scientist Group, the Chinese Academy of Sciences (LAMOST), 
 Los Alamos National Laboratory, the Max-Planck-Institute for Astronomy (MPIA), the Max-Planck-Institute 
 for Astrophysics (MPA), New Mexico State University, Ohio State University, University of Pittsburgh, 
 University of Portsmouth, Princeton University, the United States Naval Observatory, and the University 
 of Washington.
 \end{acknowledgements}

\end{twocolumn}
\end{document}